%% file: main.tex
\documentclass[11pt, letter]{article}

\input{preamble}

\begin{document}



\title{On Local Distributions in Graph Signal Processing}


\author{T.~Mitchell~Roddenberry,
  Fernando~Gama, \\
  Richard~G.~Baraniuk,
  Santiago~Segarra\thanks{All authors are with the Department of Electrical and Computer Engineering, Rice University, Houston, TX, USA.
    SS and TMR are supported by NSF CCF-2008555.}}

\date{February 2022}

\maketitle


\begin{abstract}%
Graph filtering is the cornerstone operation in graph signal processing (GSP).
Thus, understanding it is key in developing potent GSP methods.
Graph filters are local and distributed linear operations, whose output depends only on the local neighborhood of each node.
Moreover, a graph filter's output can be computed separately at each node by carrying out repeated exchanges with immediate neighbors.
Graph filters can be compactly written as polynomials of a graph shift operator (typically, a sparse matrix description of the graph).
This has led to relating the properties of the filters with the spectral properties of the corresponding matrix -- which encodes global structure of the graph.
In this work, we propose a framework that relies solely on the local distribution of the neighborhoods of a graph.
The crux of this approach is to describe graphs and graph signals in terms of a measurable space of rooted balls.
Leveraging this, we are able to seamlessly compare graphs of different sizes and coming from different models, yielding results on the convergence of spectral densities, transferability of filters across arbitrary graphs, and continuity of graph signal properties with respect to the distribution of local substructures.
\end{abstract}


\section{Introduction}\label{sec:intro}

Graph filters are a fundamental building block in graph signal processing, where cascaded applications of a graph shift operator model diffusion on the nodes of a graph~\cite{Shuman:2013, Ortega:2018}.
The analogy between filtering in discrete time and filtering on graphs has led to a fruitful research direction, with applications including power systems~\cite{Owerko:2020}, robotics~\cite{Gama:2022b}, neuroscience~\cite{Huang:2018}, and recommender systems~\cite{Ma:2016}.
Due to their typical implementation as low-degree matrix polynomials, graph filters are \emph{local operators}, where the output of a graph filter at a given node is strictly dependent on the connectivity structure and signal values on the node's local neighborhood.
This highlights an invariance property of graph filters, typically summarized by the property of \emph{permutation equivariance}.
However, the equivariance of graph filters is much stronger than not being sensitive to permutations of nodes.
If the same filter is applied to two different graphs, and two nodes within each of those graphs have identical neighborhoods, then the graph filter output at those nodes will be identical as well \cite{Gama:2020b}.
Indeed, graph filters in their usual implementation are equivariant to \emph{local substructures}, which have been shown to be of primary importance in real-world networks~\cite{Milo:2002}, often leading to useful properties such as scale invariance and robustness~\cite{Tzouanas:2021}.

The importance of local substructures is an incipient development in the literature.
This view has been considered before in the graph wavelet literature~\cite{Hammond:2011, Shuman:2015, Roddenberry:2021}, where wavelet atoms are constructed by applying graph filters to impulse functions on each node in the graph, yielding a dictionary of atoms with known spectral properties that also exhibit (approximate) spatial localization.
As a representational tool for graphs predating the development of graph signal processing, graph kernel methods have put forth the idea of graphs as bags of motifs~\cite{Kriege:2020}, such as in the paper on \emph{graphlets}~\cite{Shervashidze:2009}.
Works such as \cite{Morris:2016, Togninalli:2020} have considered the extension of graph kernels for graphs with continuous labels, typically via a neighborhood aggregation or discretization approach.
In \cite{deHaan:2020}, equivariance to local structures is proposed as a more useful invariant for graph neural networks, as opposed to permutation equivariance.
For instance, \cite{Bevilacqua:2021} considers neural networks for graph classification that act on small subgraphs over an entire graph, and \cite{Zhang:2021} considers how to design graph neural networks that can recognize structures more expressive than those of the Weisfeiler-Lehman test.
The treatment of graphs as distributions of ego-networks in \cite{Zhu:2021} was used to devise a novel loss function for training of graph neural networks based on the maximization of mutual information.
Additionally, local structures at each node need not be deterministic, as \cite{Maehara:2021} considers random walk features and defines the notion of an estimable parameter under such a model.

In this work, we develop machinery for reasoning about basic notions in graph signal processing, with the primacy of local substructures in mind.
After reviewing basic definitions for graph signal processing in \cref{sec:gsp}, we make the following theoretical contributions:
\begin{enumerate}
\item We introduce rooted graphs and rooted graph filters as vehicles for describing localized graph filters. In doing so, we develop a measure-theoretic view of graph signal processing that considers distributions of rooted balls and signals supported on them (\cref{sec:local-gsp}).
\item Within the proposed framework, we illustrate how convergence of Fourier spectra of graph signals can be understood in terms of weak convergence of measures (\cref{sec:spectra}).
\item We apply the proposed framework to yield a principled understanding of the transferability of graph summaries via integral probability metrics (\cref{sec:transfer}).
\item To highlight the flexibility of the proposed approach, we extend the relationship between distributions of local graph structures and the Fourier spectra of graph signals to weighted graphs (\cref{sec:weighted}).
\item We identify graphings and signals supported on them as the appropriate limiting objects in the proposed framework, and develop basic notions of graphing signal processing. We prove that the proposed framework applies directly to graphing signal processing in a natural way (\cref{sec:graphing}), yielding a suitable spectral theory.
\end{enumerate}


\section{Graph Signal Processing}\label{sec:gsp}

Consider an undirected graph $\gra{g} = (\set{v},\set{e})$ where $\set{v}$ is the set of nodes and $\set{e} \subseteq \set{v} \times \set{v}$ is the set of edges.
Since the graph is undirected, it holds that if edge $(u,v) \in \set{e}$ for $u,v \in \set{v}$, then $(v,u) \in \set{e}$.
We further assume the graph to have no self-loops, \ie, $(v,v) \notin \set{e}$ for all $v \in \set{v}$, and to be unweighted. An extension to weighted graphs can be found in \cref{sec:weighted}.

The neighborhood $\set{N}(\set{v}')$ of a given collection of nodes $\set{v}'\subseteq\set{v}$ is defined as
\begin{equation} \label{eq:neighborhood}
  \set{N}(\set{v}')=\{u\in\set{v}:(u,v)\in\set{e}\text{ for some }v\in\set{v}'\}\cup\set{v}'.
\end{equation}
That is, the neighborhood of a collection of nodes is that set of nodes $\set{v}'$ as well as those nodes immediately connected to it.
The $k$-hop neighborhood $\set{n}^{k}(\set{v}')$ can then be conveniently defined in a recursive manner as $\set{n}^{k}(\set{v}') = \set{n} ( \set{n}^{k-1}(\set{v}'))$ for integers $k \geq 1$ and with $\set{n}^{0}(\set{v}') = \set{v}'$.
Note that the $k$-hop neighborhood of a singleton $\set{v}' = \{v\}$ is denoted as $\set{N}^{k}(v)$, and that its degree can be easily computed as $\deg(v)=|\set{N}(v)|-1$.

Graph signals can be associated with a graph structure and are defined as a map $\vect{x}:\set{v} \to \mathbb{R}$ between the node set $\set{v}$ and the real numbers $\mathbb{R}$.
That is, a graph signal simply attaches a single real number $\vectidx{x}{v} \in \mathbb{R}$ to each node of the graph $v \in \set{v}$.
For a given graph $\gra{G} = (\set{V}, \set{E})$, we can define the space of graph signals as $\mathbb{X}(\gra{G}) = \{\vect{x}:\set{V} \to \mathbb{R}\}$.

A graph shift operator (GSO) $\mtx{S}$ can also be associated with the graph structure $\gra{G}$ as a means of relating graph signals explicitly with the underlying graph support.
More precisely, the GSO is defined as a linear operator between graph signals $\mtx{S}: \mathbb{X}(\gra{G}) \to \mathbb{X}(\gra{G})$ such that the output graph signal $\vect{y} = \mtx{S} \vect{x}$ is computed as
\begin{equation} \label{eq:graphShift}
    \vectidx{y}{v} = \sum_{u \in \set{N}(v)} \mtxidx{S}{vu} \vectidx{x}{u}.
\end{equation}
The operation \eqref{eq:graphShift} shifts the signal around the graph and is analogous to the time-shift in discrete-time signal processing.
Note that the GSO can be completely characterized by specifying the adjacency rule that assigns the values $\mtxidx{S}{vu}$ for every $(u,v) \in \set{e}$.
Examples of GSOs that have found widespread use in the literature include the adjacency matrix, the Laplacian matrix, the Markov transition matrix, and their normalized counterparts.
While it is technically possible to design adjacency rules that lead to arbitrary values of $\mtxidx{S}{vu}$, we only consider rules that are determined exclusively by the combinatorial structure of the graph.

A graph filter can then be built from the GSO.
More specifically, given a collection of $(K+1)$ scalar coefficients $\{h_{k}\}_{k=0}^{K}$, a $K$-tap graph filter $\filt{H}(\mtx{S}):\mathbb{X}(\gra{G}) \to \mathbb{X}(\gra{G})$ is defined as the linear mapping between two graph signals $\vect{y} = \filt{H}(\mtx{S}) \vect{x}$ given by
\begin{equation} \label{eq:graphFilter}
    \filt{h}(\mtx{S}) = \sum_{k=0}^{K} h_{k} \mtx{S}^{k}
\end{equation}
where $\mtx{S}^{k}$ denotes $k$ repeated applications of the GSO (see~\eqref{eq:graphShift}) to the input signal $\vect{x}$.
The operator $\mtx{S}^0$ is understood to be the identity map on $\mathbb{X}(\gra{G})$.

Graph filters are linear and local operators.
They are linear in the input graph signal $\vect{x}$.
They are local in that they only require information up to the $K$-hop neighborhood.
More specifically, the output of the graph filter at each node $\vectidx{y}{v}$ can be computed by $K$ repeated exchanges of information with one-hop neighbors $\set{N}(v)$, thus collecting the signal values contained in up to the $K$-hop neighborhood $\set{N}^{K}(v)$.
Each node can compute their output separately from the rest.
The key observation is that the nodes need not know the global structure of the whole graph $\gra{G}$ but only their local neighborhood.
Thus, graph filters can be analyzed and understood entirely from looking at this local neighborhood structure.


\section{A Local Framework for GSP}\label{sec:local-gsp}

%
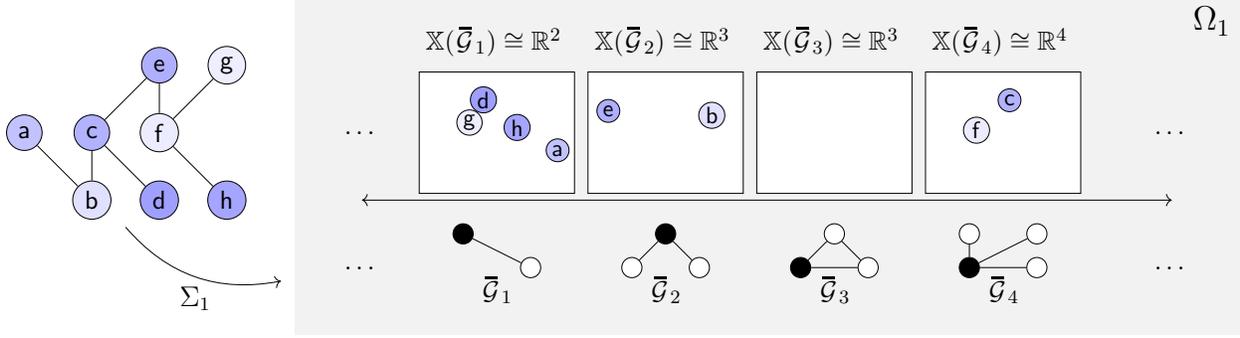
\begin{figure}
  \centering
  \resizebox{\linewidth}{!}{\input{figs/motifspace.tikz}}
  \caption{%
    The space $\Omega_K$ of rooted $K$-balls and signals, constructed via the disjoint union of signal spaces enumerated by rooted $K$-balls.
    \textbf{(Left)} A graph on nodes $\{a,b,\ldots,h\}$, with graph signal denoted by the node coloring.
    \textbf{(Right)} The space $\Omega_1$, and the distibution $\mu$ of points obtained via the pushforward of $\Sigma_1$.
    Observe that nodes $\{a,d,g,h\}$ all have isomorphic rooted $1$-balls, and thus all correspond to points in $\mathbb{X}(\rgra{G}_1)$ (depicted here in random positions within Euclidean space).
    Similarly, nodes $\{b,e\}$ are mapped to points in $\mathbb{X}(\rgra{G}_2)$, and $\{c,f\}$ are mapped to points in $\mathbb{X}(\rgra{G}_4)$.
    Since the given graph is triangle-free, no points are mapped to $\mathbb{X}(\rgra{G}_3)$.
  }
  \label{fig:motifspace}
\end{figure}
%

The local nature of graph filters calls for a local framework to analyze their effects.
Towards this end, define a rooted $K$-ball $\rgra{G}_{K}(r) = (\rset{v}_{K}, \rset{e}_{K}, r)$ as a graph with a root $r \in \rset{V}_{K}$ such that all nodes are at most $K$-hops away from the root, \ie, $\rset{v}_{K} = \set{N}^{K}(r)$.
Note that, since the rooted $K$-ball $\rgra{G}_{K}(r)$ is a graph, the notions of graph shift operator and graph signal are immediately well-defined.
We denote these as $\rmtx{S}_{K}(r)$ and $\rvect{x}_{K}(r)$, respectively.
For ease of exposition, the specification of the root node will be dropped, unless necessary to avoid confusion.

Of particular interest is the case of rooted $K$-balls obtained as the induced subgraph of the $K$-hop neighborhood of a node.
More specifically, given a graph $\gra{g} = (\set{v}, \set{e})$, a rooted $K$-ball $\rgra{g}_{K}$ can be constructed by selecting a node $r \in \set{v}$, setting $\rset{V}_{K} \equiv \set{N}^{K}(r)$ and $\rset{E}_{K} \equiv \set{e} \cap \set{N}^{K}(r) \times \set{N}^{K}(r)$.
In this context, the graph signal $\rvect{x}_{K}$ corresponding to the rooted $K$-ball can be tied to the graph signal $\vect{x}$ defined on $\gra{G}$ by setting $[\rvect{x}_{K}]_{v} = \vectidx{x}{v}\ \text{for all}\ v \in \set{N}^{K}(r)$.

We can now formalize the notion that a graph filter only relies on local information.

%
\begin{proposition}\label{prop:local-filtering}
Let $\gra{g}=(\set{v},\set{e})$ be a graph with graph signal $\vect{x}$ and a GSO $\mtx{S}$.
For a node $r\in\set{v}$, denote the corresponding rooted $K$-ball as $\rgra{G}_{K}$, with graph signal $\rvect{x}_{K}$ and GSO $\rmtx{S}_{K}$, respectively.
Then, for any $K$-tap filter $\filt{H}$, the following equality holds
\begin{equation}\label{eq:local-filtering}
    \big[ \filt{H}( \mtx{S} ) \vect{x} \big]_{r} =
    \big[ \filt{H}( \rmtx{S}_{K} ) \rvect{x}_{K} \big]_{r}.
\end{equation}
\end{proposition}
%
\noindent
We note that for \eqref{eq:local-filtering} to hold, the adjacency rule used to construct $\mtx{S}$ and $\rmtx{S}_{K}$ has to be the same, and it is assumed to depend only on the combinatorial structure of the graphs; see \cref{sec:gsp}.

\Cref{prop:local-filtering} formalizes the well-known fact that a $K$-tap graph filter evaluated at a node only depends on the subgraph of $K$-hops centered at that node (\ie, the ego-network of the node~\cite{Leskovec:2012}).
In this sense, it is not necessary to understand how a graph filter behaves on the global structure of a graph.
Rather, one only needs to understand the behavior of a graph filter locally, in particular on the rooted $K$-balls of the graph.
This naturally leads to a weaker form of the property of permutation equivariance, discussed next.

For two graphs $\gra{G}=(\set{v},\set{e})$ and $\gra{G}'=(\set{v}',\set{e}')$, we say that a map $\phi:\set{v}\to\set{v}'$ is a $K$-morphism if it preserves the structure of rooted $K$-balls for all nodes in $\set{v}$.
The structure of two rooted $K$-balls $\rgra{G}_{K} = (\rset{V}_{K}, \rset{E}_{K}, r)$ and $\rgra{G}'_{K} = (\rset{V}'_{K}, \rset{E}'_{K}, r')$ with signals $\rvect{x}_{K}$ and $\rvect{x}_{K}'$, respectively, is preserved, if there exists an isomorphism $\psi: \set{V}_{K} \to \set{V}'_{K}$ such that $\psi(r) = r'$, $(\psi(u),\psi(v)) \in \set{E}'_{K}$ if and only if $(u,v) \in \set{e}_{K}$, and $[\rvect{x}_{K}]_{v} = [\rvect{x}_{K}']_{\psi(v)}$ for all $v \in \rset{V}_{K}$.
Then, a $K$-morphism $\phi:\set{V} \to \set{V}'$ between two graphs $\gra{G}=(\set{v},\set{e})$ and $\gra{G}'=(\set{v}',\set{e}')$ with associated signals $\vect{x}$ and $\vect{x}'$, respectively, satisfies that the rooted $K$-balls $(\rgra{G}_K(v),\rvect{x}_K(v))$ and $(\rgra{G}_K'(\phi(v)),\rvect{x}_K'(\phi(v)))$ are isomorphic for all nodes $v \in \set{V}$.
Note that $K$-morphisms are not necessarily invertible, nor are they necessarily surjective.

Graph filters are invariant under $K$-morphisms, as it follows from \cref{prop:local-filtering}, where we see that the same filter applied to two different graphs yields the same result if the local structures are the same.

%
\begin{corollary} \label{cor:local-isomorphism}
Consider two graphs $\gra{G} = (\set{V}, \set{E})$ and $\gra{G}'=(\set{V}', \set{E}')$ with corresponding signals $\vect{x}$ and $\vect{x}'$.
Denote by $\rgra{G}_{K}(v)$ the $K$-rooted ball for root $v \in \set{V}$, and by $\rmtx{S}_{K}(v)$ and $\rvect{x}_{K}(v)$ the corresponding GSO and graph signal, respectively.
If there exists a $K$-morphism $\phi: \set{V} \to \set{V}'$, then for any $K$-hop filter $\filt{h}$ it holds that
\begin{equation}
    \Big[ \filt{H}\Big( \rmtx{S}_{K}\big(v \big) \Big) \rvect{x}_{K}\big(v\big) \Big]_{v} =     \Big[ \filt{H}\Big( \rmtx{S}_{K}\big( \phi(v) \big) \Big) \rvect{x}_{K}\big(\phi(v)\big) \Big]_{\phi(v)}
\end{equation}
for all $v \in \set{V}$.
\end{corollary}
%

Note that \cref{cor:local-isomorphism} is a generalization of the permutation equivariance property that graph filters have~\cite{Gama:2020b}.
More generally, \cref{cor:local-isomorphism} opens up ways to compare the performance of a fixed graph filter across two different graphs with different associated graph signals.
It would be expected then, that if two graphs have similar distributions of rooted $K$-balls with associated signal, then a fixed graph filter acting on one should yield similar results on the other.
We formalize these notions in the ensuing discussion.


\subsection{Distributions of Rooted Balls} \label{sec:local-gsp:rdistro}

\cref{prop:local-filtering} states that the output of a $K$-tap graph filter at each node depends only on the rooted $K$-ball at that node and the corresponding graph signal values.
Therefore, to characterize the effect of graph filtering, it suffices to characterize the \emph{distribution} of rooted $K$-balls of a graph.
Towards this end, we will construct a sample space, $\sigma$-field, and a corresponding measure to describe this distribution.

Denote by $\rgra{g}_{k}$ a rooted $k$-ball and by $\mathbb{X}(\rgra{g}_{k})$ its corresponding signal space.
Consider now the space constructed by the disjoint union of all signal spaces on all possible rooted $k$-balls (up to isomorphism) with $k \leq K$, given by%
\footnote{In this context, each signal space $\mathbb{X}(\rgra{g}_{k})$ is considered modulo the action of the automorphism group of $\rgra{g}_k$. This is clarified in \cref{app:automorphism}.}
\begin{equation}\label{eq:disjoint-union}
    \Omega_K = \coprod_{\rgra{g}_{k}:k\leq K}\mathbb{X}(\rgra{g}_{k}).
\end{equation}

The construction of the space $\Omega_{K}$ is illustrated in~\cref{fig:motifspace}.
Essentially, one considers a different space $\mathbb{X}(\rgra{G})$ (typically, $|\rset{V}|$-dimensional vectors $\rvect{x}$) for each possible rooted $K$-ball $\rgra{G}$, and then puts them all into a common space where each original space is isolated.
An element $\omega \in \Omega_{K}$ is described by both the rooted $K$-ball and the graph signal space it indexes, \ie, $\omega = (\rgra{G}, \mathbb{X}(\rgra{G}))$.
Making the identification $\mathbb{X}(\rgra{G}) \equiv \mathbb{R}^{|\rset{V}|}$, the point $\omega$ can be thought as a  $|\rset{V}|$-dimensional vector associated with the rooted $K$-ball $\rgra{G}$.
Note that, while it may happen that $\mathbb{X}(\rgra{G}) \cong \mathbb{X}(\rgra{G}')$ because $|\rset{V}| = |\rset{V}'|$, these are different component spaces in the disjoint union $\Omega_{K}$ as they are indexed by different rooted balls $\rgra{G}$ and $\rgra{G}'$, respectively.

To define the corresponding $\sigma$-field $\sigma(\Omega_{K})$ associated with $\Omega_{K}$, note that the disjoint union of graph signal spaces (\ie, Euclidean spaces) has a natural topology defined over it, namely the disjoint union topology.
Thus, the Borel $\sigma$-field $\sigma(\Omega_{K})$ generated from the topology of $\Omega_{K}$ is a proper $\sigma$-field.
Intuitively, measurable sets in $\Omega_{K}$ are constructed by selecting some rooted $K$-ball $\rgra{G}$ and a Borel set in the space of graph signals $\mathbb{X}(\rgra{g})$, then considering the corresponding set in $\Omega_K$.

A measure $\mu$ can be properly defined now that the $\sigma$-field associated with $\Omega_{K}$ has been constructed.
Let $\Sigma_{K} : \set{v} \to \Omega_{K}$ be a map that takes a node $v \in \set{v}$ and returns the corresponding rooted $K$-ball and signal, \ie,
\begin{equation}\label{eq:sampling-map}
    \Sigma_K(v) = \big(\rgra{G}_K(v),\rvect{x}_K(v) \big).
\end{equation}
Note that $\Sigma_{K}(v) \in \Omega_{K}$.
Then, for a Borel set $\set{A} \in \sigma(\Omega_{K})$, the measure $\mu: \sigma(\Omega_{K}) \to [0,1]$ is defined as
\begin{equation}\label{eq:pushforward}
    \mu(\set{A}) = \frac{1}{|\set{V}|} \big| \big\{ v \in \set{V} : \Sigma_{K}(v) \in \set{A} \big\} \big|.
\end{equation}
Recall that each element in $\Omega_K$ can be thought of as a rooted $K$-ball with an associated signal.
Formally, $\mu$ is the pushforward of the uniform measure on the nodes of $\gra{g}$ by the sampling map $\Sigma_{K}$, which can be denoted as $\mu=(\Sigma_K)_*(\gra{g},\vect{x})$~\cite{Tao:2011}.
We observe that any measure on the nodes of the graph can be used to replace the RHS of \eqref{eq:pushforward}.

The space $\Omega_K$ serves as a natural domain in which to characterize $K$-tap graph filters.
Indeed, \cref{prop:local-filtering} indicates that the behavior of a $K$-tap graph filter can be fully characterized by its behavior over $\Omega_K$.
That is to say, any given $K$-tap graph filter $\filt{H}$ has a corresponding map $\rfilt{H}:\Omega_K\to\mathbb{R}$ such that for any graph $\gra{G}=(\set{v},\set{e})$ with GSO $\mtx{S}$ and signal $\vect{x}$, \eqref{eq:local-filtering} can be rewritten as:
\begin{equation}\label{eq:filter-descends}
  \rfilt{H}(\Sigma_K(r)) = [\filt{H}(\mtx{S})\vect{x}]_r.
\end{equation}
To use the terminology of quotient spaces~\cite{Lee:2013}, \eqref{eq:filter-descends} indicates that a $K$-tap graph filter passes, or \emph{descends}, to a unique map on $\Omega_K$ via the map $\Sigma_K$.
Thus, in the context of \cref{prop:local-filtering} and graph filtering, it is sufficient to characterize local operations over $\Omega_K$, rather than with respect to the global structure of a given graph $\gra{G}$.

Now that the local framework based on rooted balls has been introduced, we proceed to show how it can be leveraged to obtain theoretical results that provide insight into the inner workings of graph signal processing. In \cref{sec:spectra} we discuss how to understand power spectrum densities, while in \cref{sec:transfer} we discuss transferability of graph filters across graphs of different size.

%
\begin{remark}[Distributions of edge sets]
In~\cite{Ji:2021}, the authors consider a pushforward probability measure into the Fourier domain when there is an underlying probability distribution of edges in a fixed graph.
This allows for the modeling of graph signals under uncertainty on the edge set.
Such a distribution of edges can be incorporated into the distribution $\mu$ via the pushforward map $(\Sigma_K)_*$, allowing for a description of the distribution of rooted balls in a graph under a probability distribution on the edges.
\end{remark}
%

%
\begin{remark}[Rooted subtrees]
Many authors have considered graph processing methods that are equivariant to local subtree structures~\cite{Zhang:2021}, such as architectures whose computation resembles the $1$-Weisfeiler-Lehman test\cite{Kriege:2020}.
In the proposed framework, these subtree structures are strictly ``less expressive'' than rooted ball structures, in the sense that the rooted subtree of depth $K$ centered at a given node can be determined completely by the rooted $K$-ball centered at that node.
However, the rooted $K$-ball cannot be determined by the rooted subtree of depth $K$, as evidenced by the inability of the $1$-Weisfeiler-Lehman test to distinguish certain structures~\cite{Shervashidze:2011}.
With this in mind, the proposed framework could certainly be applied where the space $\Omega_K$ is constructed with rooted subtrees, rather than rooted balls.
However, given that these rooted balls are more expressive than rooted subtrees~\cite{Zhang:2021}, we state all proceeding results using the rooted ball construction.
\end{remark}
%


\section{Power Spectral Density}\label{sec:spectra}

One of the key ideas in graph signal processing is that of the \emph{graph Fourier transform} of a graph signal~\cite{Ortega:2018}.
For a graph $\gra{g}=(\set{v},\set{e})$ on $n$ nodes, consider the GSO $\mtx{S}$ to be the graph Laplacian, which is a positive semidefinite Hermitian matrix.
Thus, it admits the following eigendecomposition
\begin{equation}
  \mtx{S} = \sum_{j=1}^n \lambda_j\vect{u}_j\vect{u}_j^\top,
\end{equation}
for eigenvalues $0=\lambda_1\leq\ldots\leq\lambda_n$ with corresponding pairwise orthogonal eigenvectors $\vect{u}_1,\ldots,\vect{u}_n$.
One can check that for any of these eigenvectors, $\langle\vect{u}_j,\mtx{S}\vect{u}_j\rangle=\lambda_j$.
The graph Laplacian induces a quadratic form on a graph signal that measures a useful notion of smoothness~\cite{Ortega:2018}.
For a given graph signal $\vect{x}$ on $\gra{g}$, the quadratic form can be shown to be equal to
\begin{equation}\label{eq:smoothness}
  \langle\vect{x},\mtx{S}\vect{x}\rangle = \sum_{v\in\set{v}}\sum_{u\in \nbhd(v)}\left(\vectidx{x}{v}-\vectidx{x}{u}\right)^2.
\end{equation}
That is, the quadratic form of the graph Laplacian measures the sum of the squared differences of the graph signal at neighboring pairs of nodes.
A signal is called \emph{smooth} if the quadratic form takes a small value relative to its norm.
For instance, the eigenvectors of $\mtx{S}$ with small corresponding eigenvalues are smooth signals.

We note that this notion of smoothness can be extended to normalized versions of the graph Laplacian in a similar fashion. Furthermore, a notion of smoothness can also be defined using the adjacency matrix~\cite{Gama:2019}.
In any case, for ease of exposition and conceptual simplicity, we assume that $\mtx{S}$ is the graph Laplacian throughout this section, but we remark that the results derived herein are also valid for any other GSO built from adjacency rules that rely on the combinatorial structure of the graph.

Since the eigenvectors $\vect{u}_1,\ldots,\vect{u}_n$ form an orthonormal basis for $\mathbb{R}^n$, any graph signal $\vect{x}$ admits a unique representation as a linear combination of the eigenvectors of $\mtx{S}$.
That is, for any graph signal $\vect{x}$, there exists a set of coefficients $\{\tilde{x}_j\}_{j=1}^n$ such that the following holds:
\begin{align}
  \vect{x} &= \sum_{j=1}^n\tilde{x}_j\vect{u}_j \label{eq:graph-fourier-transform}\\
  \langle\vect{x},\mtx{S}\vect{x}\rangle &= \sum_{j=1}^n\lambda_j\tilde{x}_j^2.
\end{align}
Thus, the representation of a graph signal as a weighted sum of the eigenvectors can be used to conveniently compute the quadratic form in terms of the spectrum.
In an analogy to complex exponential functions being the eigenfunctions of the Laplace operator in discrete-time signal processing, we call the representation of the graph signal $\vect{x}$ by the coefficients $\tilde{x}_j$ the \emph{graph Fourier transform} (GFT).

When coupled with their corresponding eigenvalues $\lambda_j$, the coefficients $\tilde{x}_j$ can be used to describe the distribution of energy in a graph signal, in a way concordant with the notion of smoothness described by the Laplacian quadratic form~\eqref{eq:smoothness}.
To capture this, define the \emph{normalized power spectral distribution} of a graph signal .
Given a graph signal $\vect{x}$ with Fourier coefficients $\tilde{x}_j$ and corresponding Laplacian eigenvalues $0=\lambda_1\leq\ldots\leq\lambda_n$, define
\begin{equation}\label{eq:graph-psd}
  \begin{aligned}
    P_{\vect{x}}:\mathbb{R}&\to\mathbb{R}^+ \\
    \lambda&\mapsto\frac{1}{n}\sum_{j:\lambda_j\leq\lambda}\tilde{x}_j^2.
  \end{aligned}
\end{equation}
One can easily see that $P_{\vect{x}}$ is a monotone, right-continuous function, with $P_{\vect{x}}(\lambda)=0$ for $\lambda<0$, and at most $n$ discontinuities (one at each eigenvalue $\lambda_j$).
Moreover, there is a convenient expression for the moments of $P_{\vect{x}}$:
\begin{equation}\label{eq:psd-moments}
  \moment{m}_K(\vect{x}) \coloneqq \int_{\mathbb{R}}\lambda^K\drv P_{\vect{x}}(\lambda) = \frac{1}{n}\langle\vect{x},\mtx{S}^K\vect{x}\rangle.
\end{equation}
That is to say, the moments of the normalized power spectral distribution are given by scaled quadratic forms of powers of the GSO.
Observe that the quadratic form~\eqref{eq:smoothness} can be expressed as $\langle\vect{x},\mtx{S}\vect{x}\rangle=n\cdot\moment{m}_1(\vect{x})$.
Indeed, the moments of the graph Fourier transform are an essential quantity in the design of low-order graph filters~\cite{Segarra:2017b}.
This points to a very useful idea: the powers of the GSO are the most primitive types of graph filters, so the proposed local framework provides a path for understanding these moments in terms of functions on the space of distributions of rooted balls.

To see this, let us examine \eqref{eq:psd-moments}.
Notice that
\begin{equation}
  \frac{1}{n}\langle\vect{x},\mtx{S}^K\vect{x}\rangle = \frac{1}{n}\sum_{v\in\set{v}}\vectidx{x}{v}[\mtx{S}^K\vect{x}]_v.
\end{equation}
As before, let $\rmtx{S}_{K}(v)$ be the GSO of the $K$-ball rooted at node $v$.
By \cref{prop:local-filtering}, the above equation can be written as
\begin{equation}
  \frac{1}{n}\sum_{v\in\set{v}}\vectidx{x}{v}[\mtx{S}^K\vect{x}]_v =
  \frac{1}{n}\sum_{v\in\set{v}}[\rvect{x}_K]_{v}[\rmtx{S}_{K}^{K}(v)\rvect{x}_K(v)]_v.
\end{equation}
That is, the moments of the normalized power spectral distribution of a graph signal can be written as an average of operations that resemble local filtering.
In the proposed local framework, this has the following interpretation.

%
\begin{proposition}\label{prop:moments-expectations}
  Let $\gra{g}$ be a graph and $\vect{x}$ be a graph signal.
  For a given $K\geq 0$, define the map
  \begin{align}
    \rmoment{m}_K:\Omega_K&\to\mathbb{R} \\
    ((\set{v},\set{e},r),\rvect{y})&\mapsto\rvectidx{y}{r}[\rmtx{S}_K^K\rvect{y}]_r,
  \end{align}
  %
  where $\rmtx{S}_K$ denotes the Laplacian of the rooted $K$-ball.

  Letting $\mu$ be the probability distribution on $\Omega_K$ determined by $(\set{g},\vect{x})$, the $K^{th}$ moment of the normalized power spectral distribution $P_{\vect{x}}$ is given by the equation
  \begin{equation}\label{eq:moment-expectation}
    \moment{m}_K(\vect{x})
    = \int_{\Omega_K}\rmoment{m}_K(\omega)\drv\mu(\omega)
    = \mathbb{E}_\mu [ \rmoment{m}_K].
  \end{equation}
\end{proposition}
%

Having reduced the moments of the normalized power spectral distribution to an average over the distribution of rooted balls, we can now reason about the convergence of graph Fourier distributions in terms of weak convergence of measure for graphs with bounded degree $D$.
Although the limit of dense graphs has been considered before via graphon models~\cite{Avella:2018, Ruiz:2021a}, this cannot capture the behavior of bounded degree graphs, as all infinite sequences of growing graphs of bounded degree converge to the zero graphon~\cite{Borgs:2018}.
However, in the case of sparse graphs, understanding the descension of maps to the space $\Omega_K$ is a feasible approach, with the following compactness property.
\begin{lemma}\label{lemma:compactness}
  For given integers $K\geq 0,D\geq 1$ and any collection of graphs in $\graphsd{D}$ with uniformly bounded signals, there exists a compact subspace $\Gamma\subset\Omega_K$ such that the support of the probability measures corresponding to the graphs/graph signals is contained in $\Gamma$.
\end{lemma}
The proof of \cref{lemma:compactness} can be found in \cref{app:proof:compactness}.
The constraint of a graph having bounded degree, for instance coming from physical constraints in a real-world system, corresponds to a compact description in the space $\Omega_K$.
From this, the following convergence result holds.

%
\begin{theorem}\label{thm:psd-convergence}
  Let integers $K\geq 0, D\geq 1$ be given.
  Let $\{(\gra{g}_j,\vect{x}_j)\}_{j=1}^\infty$ be a sequence of graphs and graph signals such that each $\gra{g}_j\in\graphsd{D}$ and the collection of signals $\vect{x}_j$ is uniformly bounded.
  Let $\mu_j$ be the probability measure on $\Omega_K$ associated with $(\gra{g}_j,\vect{x}_j)$ for each $j$, and $P_j$ the corresponding normalized power spectral distribution function.

  If the measures $\mu_j$ converge weakly, then the $K$th moments of the normalized power spectral distribution functions converge.
  Moreover, if weak convergence of measure holds for all $K\geq 0$, then the normalized power spectral distribution functions converge weakly.
\end{theorem}
The proof of \cref{thm:psd-convergence} can be found in \cref{app:proof:psd-convergence}.
\Cref{thm:psd-convergence} casts the power spectral density of a graph signal in terms of the distribution of local structures in the graph.
In particular, we treat the power spectral density as a function from graphs with graph signals to Borel measures on a compact subset of the real line, then prove that this function is continuous with respect to the distribution of rooted balls in the graph, where continuity is understood to be with respect to the weak topology for both the distribution in $\Omega_K$ and the power distribution of the signal.
Although the frequency domain representation of a graph signal is typically understood to be a representation in terms of the \emph{global} graph structure, owing to the global support of the graph Laplacian's eigenvectors, this results indicates that it is in essence still subject to fundamentally local phenomena, namely the rooted balls in the graph.

This characterization of the power spectral distribution in terms of the distribution of rooted balls also reflects the local nature of graph filtering.
Indeed, as discussed in~\cite{Segarra:2017b}, the moments of the power spectral distribution are key in the understanding of graph filters.
More broadly, when designing a graph filter, we often aim to construct a \emph{matched filter} in the frequency domain, so that the response of the filter aligns with that of the signal, subject to some noise model (additive white Gaussian noise, for instance).
Additionally, constraints such as a filter having $K$ taps reduces the space of possible frequency responses to degree $K$ polynomials in the frequency domain, so that the moments become the basic values of interest.
This is an immediate consequence of graph filters admitting a distributed implementation~\cite{Segarra:2017b}, so that the expression of a graph filter's performance in terms of the power spectrum of the signal is actually a distillation of the distribution of rooted balls in $\Omega_K$.

We have shown how connecting the moments of the power spectral distribution to the space $\Omega_K$ yields insights about how features of graphs and graph signals are dependent only on localized information.
In particular, a function $\rmoment{m}_K:\Omega_K\to\mathbb{R}$ was constructed whose expected value under the pushforward measure of a graph computes the moments of the power spectral distribution.
This machinery can be immediately extended to other functions on graphs that act locally, allowing us to compare the behavior of a given function on two graphs in terms of their distributions of rooted balls.


\section{Transferability}\label{sec:transfer}

Let $J$ be a function mapping a graph and its corresponding graph signal to a real number.
This function is typically known as a \emph{graph summary} and examples include the graph power spectral distribution discussed in the previous section, as well as the average node degree or node centrality values~\cite{Segarra:2015}, the clustering coefficient~\cite{Watts:1998}, or the conductance~\cite{Bollobas:1998}.
We can also consider \emph{cost functions} for a given task as graph summaries, since they ultimately take a graph and its signal as input, and output a real number \cite{Gama:2020a, Gama:2022a}.
In this context, we want to study how the function $J$ changes across two different graphs with different graph signals $(\gra{G}_{1},\vect{x}_{1})$ and $(\gra{G}_{2},\vect{x}_{2})$.

To study the transference of the graph summary $J$ across different graphs and graph signals, let us assume that there exists a function $\rooted{J}:\Omega_{K} \to \mathbb{R}$ such that for any pair $(\gra{g},\vect{x})$ whose associated probability distribution on $\Omega_K$ is denoted by $\mu$, we have $J(\gra{g},\vect{x})=\mathbb{E}_{\mu}[\rooted{J}]$.
This assumption is not too severe, essentially saying that the graph summary is an average over a function on the nodes, with the value at each node determined strictly by the structure of its local neighborhood.
Then, consider the \emph{transfer equation}
\begin{equation}\label{eq:transfer-equation}
    \left|J(\gra{g}_1,\vect{x}_1)-J(\gra{g}_2,\vect{x}_2)\right| = \left|\mathbb{E}_{\mu}[\rooted{J}]-\mathbb{E}_{\nu}[\rooted{J}]\right|.
\end{equation}
as a way of measuring the transferability of $J$, and where $\mu$ and $\nu$ are the distributions of rooted balls on the space $\Omega_{K}$ of the graphs $(\gra{G}_{1}, \vect{x}_{1})$ and $(\gra{G}_{2},\vect{x}_{2})$, respectively.

To understand \eqref{eq:transfer-equation} in a quantitative manner demands a stronger structure on $\Omega_K$ than the disjoint union topology.
To this end, we then endow $\Omega_K$ with the structure of a metric space, in a way that preserves the original topology.
For an arbitrary constant $C>0$, define a metric $d_C$ on $\Omega_K$ as follows.
For $\omega_1=(\rgra{g}_1,\rvect{x}_1)$ and $\omega_2=(\rgra{g}_2,\rvect{x}_2)\in\Omega_K$, put
\begin{equation}
  d_C(\omega_1,\omega_2)=
  \begin{cases}
    \min\{\|\rvect{x}_1-\rvect{x}_2\|_2,2C\} & \rgra{g}_1=\rgra{g}_2 \\
    C & \text{otherwise},
  \end{cases}
\end{equation}
where $\|\cdot\|_2$ is inherited from the identification of $\mathbb{X}(\rgra{G})$ with Euclidean space.\footnote{This metric is again understood modulo the automorphism group of the rooted ball, discussed in \cref{app:automorphism}.}
Note that $\|\cdot\|_{2}$ is well defined since both $\vect{x}_{1}$ and $\vect{x}_{2}$ are vectors in the same $|\rset{V}|$-dimensional space when $\rgra{G}_{1} = \rgra{G}_{2}$.
One can check that $d_C$ is indeed a metric on $\Omega_K$ for any $C>0$.
The metric $d_C$ is constructed in a way that preserves the \emph{local} metric structure of each signal space $\mathbb{X}(\rgra{G})$, while also endowing the whole space $\Omega_K$ with a \emph{global} metric structure, at the cost of some local distortion in order to maintain the triangle inequality, as dictated by the constant $C$.
When $C$ is chosen to be very large, the local metric structure on each rooted ball of the space is well-preserved, but these rooted balls are otherwise kept ``far apart'' from each other.
On the other hand, when $C$ is very small, the rooted balls become ``close,'' but the local metric structure on each of them is distorted (i.e. signals belonging to the same rooted ball that are beyond a distance of $2C$ are indistinguishable from signals that belong to different rooted balls).
Importantly, the metric topology on $\Omega_K$ induced by $d_C$ is equal to the disjoint union topology~\eqref{eq:disjoint-union}, so that all notions of continuity and weak convergence of measure are preserved.

Given the additional metric structure on $\Omega_K$, we can quantitatively characterize the transfer equation~\eqref{eq:transfer-equation} by comparing distributions of rooted balls on the metric space $(\Omega_K,d_C)$.
To do so, we make the following assumption on the function $\rooted{J}$.

%
\begin{assumption}\label{assump:lipschitz}
  For all rooted $K$-balls $\rgra{g}\in\rgraphsd{D}$, $\rooted{J}(\rgra{g},\rvect{x})$ is $L$-Lipschitz continuous with respect to the second argument $\rvect{x}$ over the space of bounded signals $\mathbb{X}(\rgra{g},[-1,1])$.
\end{assumption}
%
%
This is not a difficult assumption to satisfy. For instance, it is sufficient for $\rooted{J}$ to be continuously differentiable on each signal space for this to hold.
In fact, the set of Lipschitz continuous functions on a compact space is dense in the set of continuous functions with respect to the uniform norm~\cite{DeBranges:1959}.
%
%
\begin{theorem}\label{thm:transferability}
Let graphs $\gra{g}_1,\gra{g}_2\in\graphsd{D}$ with corresponding bounded signals $\vect{x}_1,\vect{x}_2$ be given.
Denote their corresponding measures in $\Omega_K$ by $\mu$ and $\nu$, respectively.
For a function $\rooted{J}:\Omega_K\to[0,1]$ that satisfies \cref{assump:lipschitz}, it holds that
\begin{equation}\label{eq:wasserstein-bound}
  \big|\mathbb{E}_{\mu}[\rooted{J}]-\mathbb{E}_{\nu}[\rooted{J}]\big|
  \leq L\cdot W_1\left(\mu,\nu;\frac{1}{L}\right),
\end{equation}
where $W_1(\mu,\nu;C)$ denotes the $1$-Wasserstein distance between $\mu$ and $\nu$ under the metric $d_C$.
\end{theorem}
%

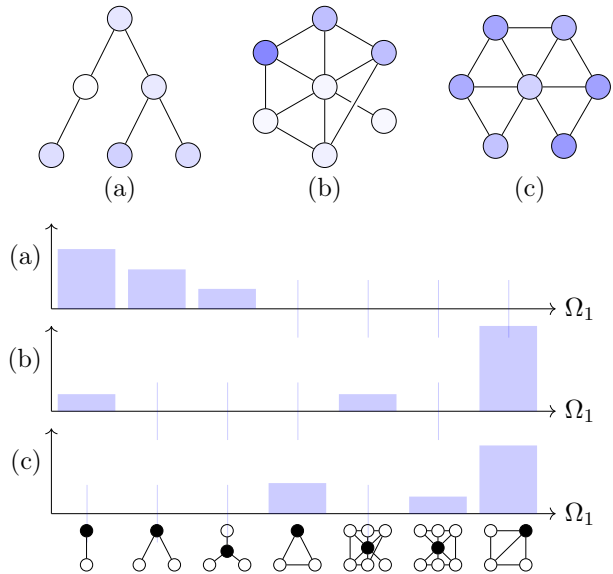
\begin{figure}
  \centering
  \resizebox{0.5\linewidth}{!}{\input{figs/wasserstein.tikz}}
  \caption{%
    The $1$-Wasserstein distance reflects structural similarity between graphs.
    \textbf{(Top)} Three graphs (a,b,c) with signals indicated by node coloring.
    \textbf{(Bottom)} Distributions of rooted $1$-balls in each graph (a,b,c), represented by histograms.
    The $x$-axis corresponds to the space $\Omega_1$, and the $y$-axis shows the density of each rooted $1$-ball in the graph.
    The $1$-Wasserstein distance, then, essentially describes the distance between histograms, subject to a metric structure on $\Omega_1$.
    For instance, graph (a) is expected to have a large $1$-Wasserstein distance from graphs (b,c), since there is minimal overlap between their histograms.
    On the other hand, graphs (b,c) have substantial overlap, so their $1$-Wasserstein distance will be smaller.
  }
  \label{fig:wasserstein}
\end{figure}

The proof of \cref{thm:transferability} can be found in \cref{app:proof:transferability}.
\Cref{thm:transferability} bounds the transfer equation~\eqref{eq:transfer-equation} in terms of the $1$-Wasserstein distance between the two distributions of rooted balls.
The appearance of the $1$-Wasserstein distance in this setting is intuitive: the dual formulation of the $1$-Wasserstein distance defines the distance between two distributions via an integral probability metric over $1$-Lipschitz functions~\cite{Sriperumbudur:2009}, yielding a transferability bound that holds for all Lipschitz functions.
We illustrate the notion of the $1$-Wasserstein distance between graphs in~\cref{fig:wasserstein}.

%
%

\subsection{Examples and Discussion}\label{sec:transfer:disc}

\Cref{thm:transferability} indicates that smoother functions $\rooted{J}$ generally transfer better than ones with large Lipschitz constant.
Some examples of graph summaries that fulfill the conditions of \cref{thm:transferability} follow.

%
\begin{parexample}[Power spectral distribution]
  For any $K\geq 0$, the spectral moment function $\moment{m}_K$ as defined in \cref{sec:spectra} descends to the expectation of the function $\rmoment{m}_K:\Omega_K\to\mathbb{R}$.
  For any rooted $K$-ball $\rgra{g}$, it holds that $\rmoment{m}_K(\rgra{g},\cdot)$ is a polynomial of the graph signal in the second argument, and can thus be shown to satisfy \cref{assump:lipschitz}.
  Thus, the difference in the spectral moments of two graphs can be bounded in terms of the Wasserstein distance of their respective distributions of rooted balls.
  Moreover, observe that for $K\geq M\geq 0$, the moment function $\rmoment{m}_M$ can be defined on $\Omega_K$ in a way that preserves the identity~\eqref{eq:moment-expectation}, so that we can compare the similarity of spectral moments between graph signals by comparing the Lipschitz constants of the functions $\rmoment{m}_K$ and $\rmoment{m}_M$ on the same space $\Omega_K$.
  Indeed, one can see that $\rmoment{m}_K$ has a larger Lipschitz constant than $\rmoment{m}_M$, so that the bound given by~\cref{thm:transferability} is tighter for the lower moments of the power spectral distribution, contingent upon a relatively small $1$-Wasserstein distance between the distributions of rooted balls.
\end{parexample}
%

%
\begin{parexample}[MSE of a graph filter]\label{exam:mse}
  Let $\filt{h}$ be a fixed $K$-tap graph filter, and let $\sigma^2>0$ be given.
  For any given graph $\gra{g}$ on $n$ nodes with associated signal $\vect{x}$ and shift operator $\mtx{S}$, put the function $J$ as
  \begin{equation}
    J(\gra{g},\vect{x}) =
    \mathbb{E}\left[\frac{1}{n}\big\|\vect{x}-\filt{h}(\mtx{S})(\vect{x}+\boldsymbol{\eta})\big\|_2^2\right],
  \end{equation}
  where the expectation is taken over the random vector $\boldsymbol{\eta}\sim\nbhd(0,\sigma^2\mathbf{I})$.
  In words, $J$ is the mean squared error when applying $\filt{h}$ to $(\gra{g},\vect{x})$ under additive white Gaussian noise.
  Define $\rooted{J}$ on rooted $K$-balls $\rgra{g}=(\set{v},\set{e},r)$ with shift operator $\rmtx{S}$ as
  \begin{equation}
    \rooted{J}(\rgra{g},\rvect{x}) =
    (\rvectidx{x}{r}-\vectidx{\filt{h}(\rmtx{S})\rvect{x}}{r})^2 +
    \sigma^2[(\filt{h}(\rmtx{S}))^2]_{r}.
  \end{equation}
  If we denote the measure on $\Omega_K$ associated with $(\gra{g},\vect{x})$ by $\mu$, one can show that
  \begin{equation}
    J(\gra{g},\vect{x}) = \mathbb{E}_{\mu}[\rooted{J}].
  \end{equation}
  Observe that $\rooted{J}(\rgra{G},\rvect{x})$ is a polynomial of the values of $\rvect{x}$ for each rooted $K$-ball $\rgra{G}$, so that it is Lipschitz continuous on each bounded signal space, thus satisfying~\cref{assump:lipschitz}.
  Moreover, the Lipschitz constant of $\rooted{J}$ over each signal space $\mathbb{X}(\rgra{G})$ can be shown to be proportional to $\|(\mtx{I}-\filt{H}(\rmtx{S}))\boldsymbol{\delta}_r\|_2$, where $\boldsymbol{\delta}_r$ is the signal taking value $1$ on the root node and $0$ elsewhere.
  That is to say, the Lipschitz constant is directly influenced by the ``spread'' of the impulse response of the filter.
  Intuitively, this indicates that while filters with large coefficients in their higher-order terms may suit a particular signal well, they may transfer across graphs poorly.
  This aligns with existing results on the stability of graph filters~\cite{Gama:2020b, Ruiz:2021b}.
\end{parexample}
%

%
\begin{parremark}[Estimation of the transferability bound]
  The computation of the $1$-Wasserstein distance in~\eqref{eq:wasserstein-bound} may prove problematic for large graphs.
  However, given the definition of the $1$-Wasserstein distance as an integral probability metric, it is not difficult to compute an emprical estimate of $W_1(\mu,\nu;C)$.
  By~\cite[Theorem~6]{Sriperumbudur:2009}, if $\{\omega_j^\mu\}_{j=1}^m$ and $\{\omega_j^\nu\}_{j=1}^n$ are {\iid} samples drawn from the respective distributions $\mu$ and $\nu$, then the $1$-Wasserstein distance between the empirical distributions can be computed via a linear program.
  To draw {\iid} samples from these distributions, one can choose a node uniformly at random (with replacement) from the node set of the underlying graph and then take the rooted $K$-ball via the map $\Sigma_K$.
  By~\cite[Corollary~10]{Sriperumbudur:2009}, the compact support of the distributions $\mu,\nu$ guarantees almost sure convergence of the empirical estimator to the true Wasserstein distance as $m,n\to\infty$.
\end{parremark}
%

%
\begin{parremark}[Tighter bounds]
In \cref{thm:transferability}, the difference in expectation between two distributions of rooted balls is characterized in terms of their Wasserstein distance, which is defined in terms of a metric imposed upon $\Omega_K$.
Given that the metric $d_C$ is defined for essentially arbitrary $C$, one can derive a tighter bound on the transfer equation, with less stringent assumptions on the function $\rooted{J}$.
First, note that for two graphs $\gra{G}_1,\gra{G}_2\in\graphsd{D}$ and bounded signals $\vect{x}_1,\vect{x}_2$, the corresponding measures $\mu,\nu$ have support in the compact subspace $\Gamma\subset\Omega_K$, where $\Gamma$ is as described in \cref{lemma:compactness}.
Because of this, any continuous function on $\Gamma$ is bounded.
For a function $\rooted{J}:\Omega_K\to\mathbb{R}$ satisfying \cref{assump:lipschitz}, define the range of $\rooted{J}$ as
\begin{equation}
  A = \sup_{\omega\in\Gamma}\rooted{J}(\omega) - \inf_{\omega\in\Gamma}\rooted{J}(\omega).
\end{equation}
In the same vein as \cref{thm:transferability}, it holds that
\begin{equation}
  \big|\mathbb{E}_{\mu}[\rooted{J}]-\mathbb{E}_{\nu}[\rooted{J}]\big|
  \leq \inf_{C\in\hintOC{0,1}}\frac{L}{C}\cdot W_1\left(\mu,\nu;\frac{AC}{L}\right).
\end{equation}
This bound illustrates an interplay between the smoothness of $\rooted{J}$, the range of $\rooted{J}$, and the regime of similarity between the two graphs.
If two graphs have similar distributions of rooted balls, then the smoothness of $\rooted{J}$ plays a stronger role in bounding transferability.
If they have highly distinct distributions of rooted balls, the range of $\rooted{J}$, as described by the value $A$, plays a stronger role in bounding transferability.
\end{parremark}
%

%
\begin{parremark}[Quantifying weak convergence]
  It is well known that in cases of bounded support, the Wasserstein distance between probability measures metrizes the weak topology on the set of probability measures on that space, \ie, the topology induced by the definition of weak convergence~\cite{Villani:2009}.
  Therefore, convergence in the $1$-Wasserstein distance is equivalent to weak convergence of measures.
  For instance, this implies via \cref{thm:psd-convergence} that the moments of the graph Fourier transform are continuous with respect to the $1$-Wasserstein distance between distributions of rooted balls.
\end{parremark}
%

%
\begin{parremark}[Extensions to graph neural networks]
  Although we have considered maps on graphs with real-valued signals, the analysis in this section is equally applicable when the nodes have other types of features such as categorical labeling.
  For example, one could consider $J$ to be the cross-entropy loss for a fixed graph neural network with $K$ layers, where the ``graph signal'' is taken to be a set of discrete input features and a ground-truth label against which the loss is evaluated \cite{Gama:2020c}.
\end{parremark}
%

%
\begin{parremark}[Wasserstein graph kernels]
  Wasserstein distances between graphs have been considered in the graph kernel literature~\cite{Togninalli:2020}, where node embeddings in Euclidean space are computed from local structures about each node, followed by a computation of the Wasserstein distance in Euclidean space.
  Given that the proposed computation for nodes with continuous attributes in this paper is smooth with respect to the node attributes, one can show that the Wasserstein distance between distributions in our case descends to theirs via a continuous map.
\end{parremark}
%

\vspace{2mm}

By endowing the space $\Omega_K$ with an appropriate metric structure, we have applied the proposed framework to studying transferability of graph summaries that descend to maps on rooted $K$-balls.
In analyzing the relationship between the Lipschitz constant of the graph summary and the distribution of rooted balls, insights have been gleaned as to what drives commonalities between graphs in this regard.
Although our focus in this section has been on fixed graph summaries, such as the loss of a fixed graph filter in~\cref{exam:mse}, these ideas can be readily adapted to problems of designing graph summaries.
For instance, the presence of the Lipschitz constant in~\cref{thm:transferability} indicates that a graph filter or graph neural network will transfer better if it is designed to be smooth.
This can be cast in the light of stability, where certain conditions on filter banks in graph neural networks can guarantee a response that is stable to structural perturbations~\cite{Gama:2020b,Ruiz:2021b}.

\section{Extension to Weighted Graphs}\label{sec:weighted}

In this section, we consider how ideas from the unweighted case (\cref{sec:spectra,sec:transfer}) can be extended to weighted graphs.
We define a \emph{weighted graph} as a graph $\gra{g}=(\set{v},\set{e})$ coupled with a weight function $w:\set{e}\to\mathbb{R}_{\geq 0}$.
For a given graph $\gra{g}$, the set of all possible weight functions on it is denoted $\mathbb{W}(\gra{g})=\{w:\set{e}\to\mathbb{R}_{\geq 0}\}$.
We use the same notation for the set of weight functions on a rooted graph: $\mathbb{W}(\rgra{g})$.

We now expand our definition of a rooted $K$-ball with a signal to include weight functions.
Define
\begin{equation}
  \Omega_K^W = \coprod_{\rgra{g}_k:k\leq K}\mathbb{W}(\rgra{g})\times\mathbb{X}(\rgra{g}).
\end{equation}
Similar to the definition of $\Omega_K$ in~\eqref{eq:disjoint-union}, the space $\Omega_K^W$ consists of all rooted $K$-balls with both edge weights and graph signals on them.
As before, a weighted graph with a signal $(\gra{g},w,\vect{x})$ yields elements of $\Omega_K^W$ via a suitably defined sampling map $\Sigma_K$, defined for nodes $v\in\set{V}$:
\begin{equation}
  \Sigma_K(v) = (\rgra{G}_K(v),\rooted{w}_K(v),\rvect{x}_K(v)),
\end{equation}
where $\rooted{w}_K(v)$ denotes the restriction of $w$ to the edges of $\rgra{G}_K(v)$.
As before, a given weighted graph with a signal defines a probability measure on $\Omega_K^W$ via the pushforward of the uniform measure by $\Sigma_K$, which we denote by $\mu=(\Sigma_K)_*(\gra{g},w,\vect{x})$.
For a weighted graph $(\gra{g},w)$, define the \emph{weighted degree} of a node $v\in\set{V}$ as follows:
\begin{equation}
  \mathrm{deg}(v) = \sum_{u\in\nbhd(v)}w(v,u).
\end{equation}
The set of weighted graphs whose nodes have weighted degree at most $D$ is denoted by $\graphswd{D}$ for real number $D>0$.

We now develop a local description of the graph Fourier transform for weighted graphs, much like in \cref{sec:spectra}.
For a graph $\gra{g}=(\set{v},\set{e})$ with weight function $w$, define the \emph{weighted graph Laplacian} as a shift operator such that, for any $u,v\in\set{v}$,
\begin{equation}
  [\mtx{S}]_{uv} =
  \begin{cases}
    \mathrm{deg}(v;w) & u=v \\
    -w(u,v) & (u,v)\in\set{e} \\
    0 & \text{otherwise}.
  \end{cases}
\end{equation}
The weighted Laplacian for a graph on $n$ nodes is positive semidefinite, and admits an eigendecomposition $\mtx{S}=\sum_{j=1}^n\lambda_j\vect{u}_j\vect{u}_j^\top$, with a Fourier representation for graph signals defined in a way analogous to~\eqref{eq:graph-fourier-transform}.
For a given graph signal $\vect{x}\in\mathbb{X}(\gra{g})$, there is a power spectral distribution associated with $\vect{x}$ via the weighted Laplacian, whose moments are given by~\eqref{eq:psd-moments}.
We characterize the properties of the power spectral distribution in terms of the distribution of rooted $K$-balls in the following theorem.

\begin{theorem}\label{thm:weighted-psd-convergence}
  Let an integer $K\geq 0$ and a real number $D>0$ be given.
  Let $\{(\gra{g}_j,w_j,\vect{x}_j)\}_{j=1}^\infty$ be a sequence of graphs, weight functions, and graph signals such that each graph is contained in $\graphswd{D}$ and the signals $\vect{x}_j$ are uniformly bounded.
  Let $\mu_j$ be the probability measure on $\Omega_K^W$ associated with $(\gra{g}_j,w_j,\vect{x}_j)$ for each $j$, and $P_j$ the corresponding normalized power spectral distribution function.

  If the measures $\mu_j$ converge weakly, then the $K^{th}$ moments of the normalized power spectral distribution functions converge.
  Moreover, if weak convergence of measure holds for all $K\geq 0$, then the normalized power spectral distribution functions converge weakly.
\end{theorem}

The proof of \cref{thm:weighted-psd-convergence} can be found in \cref{app:proof:weighted-psd-convergence}.
\Cref{thm:weighted-psd-convergence} establishes that the power spectral distribution of a graph signal on a weighted graph is continuous with respect to the weak topology of distributions of rooted $K$-balls, under boundedness assumptions.
Unlike the regime of~\cref{thm:psd-convergence}, the convergence of the power spectral distribution for weighted graphs here is not dependent on the combinatorial constraint of having bounded node degrees.
Rather, any node may have an arbitrarily large number of neighbors, as long as the total influence remains bounded.
This is a realistic assumption for many systems, where an agent may be allowed to exert influence on a large collection of other agents, but the total influence is bounded by some physical constraint, such as power consumption.

\begin{remark}[Zero-weight edges]
  In many applications, an edge with an assigned weight of zero is treated as a non-edge.
  For instance, the moment function $\rmoment{m}_K$ satisfies this condition.
  This motivates a quotient map on $\Omega_K^W$ that ``glues'' rooted balls with edge-weights that take value zero to rooted balls where those edges are not present.
  Any function that is continuous on $\Omega_K^W$ and also satisfying this condition will also be continuous under such a quotient map, so that our results still hold.
  The construction in this case is somewhat technical, so we omit it for simplicity of exposition.
\end{remark}

\section{In the Limit: Graphing Signal Processing}\label{sec:graphing}

In~\cref{sec:transfer,sec:weighted}, the (weak) continuity of the power spectral density with respect to the distribution of rooted balls was established.
Specifically, it was shown that weakly convergent sequences of distributions of rooted balls yielded weakly convergent power spectra.
Given that these distributions correspond to underlying graphs, it remains to be established precisely what objects these sequences are converging to.
In general, a weakly convergent sequence of finite graphs does not necessarily converge to a finite graph.
When studying graph limits using graphon models, the homomorphism density of motifs is of primary concern~\cite{Lovasz:2012}.
This setting is slightly different, depending on the isomorphism density of rooted graphs.
In this setting, a more appropriate model is that of a graphing.
We show that the basic ideas in the discussion so far can be transferred directly to these limiting objects.
Let us define the basic object of study for this section.

\begin{definition}\label{defn:graphing}
  A \emph{graphing} of degree $D$ is a triplet $\ging{G}=(\set{v},\set{e},\lambda)$ such that
  \begin{enumerate}
  \item $\set{V}$ is a sample space with a $\sigma$-field $\mathcal{B}$ on $\set{V}$
  \item $\lambda$ is a probability measure on $\mathcal{B}$
  \item $\set{E}\in\mathcal{B}\times\mathcal{B}$ is such that, for all $A,A'\in\mathcal{B}$, we have
  \end{enumerate}
  \begin{equation}
    \int_A\mathrm{deg}(v,A')d\lambda(v) = \int_{A'}\mathrm{deg}(v,A)d\lambda(v),
  \end{equation}
  where $\mathrm{deg}(v,A)=|\{u\in A:(u,v)\in \set{E}\}|$, with the condition that $\mathrm{deg}(v,\set{V})\leq D$ for all $v\in\set{V}$.
\end{definition}

A graphing is a way to describe a graph with a potentially uncountable number of nodes (elements of $\set{V}$), yet with bounded node degrees.
We tacitly assume that all graphings considered are of degree $D$, for some $D\geq 1$.
Unlike a graphon, which describes dense graphs with unbounded node degrees, the structures that arise from a graphing are typically sparse, and notions of graph filtering and graph shifts reduce to finite sums, rather than continuous integrals.
Given a graphing $\ging{g}=(\set{v},\set{e},\lambda)$, a \emph{graphing signal} is a map $\vect{x}:\set{V}\to\mathbb{R}$ between the nodes $\set{V}$ and the real numbers $\mathbb{R}$ such that $\vect{x}$ is a measurable function.
Endowing this with the usual vector space structure, we define the space of graphing signals as
\begin{equation}
  \mathbb{X}(\ging{g})=\{\vect{x}:\set{V}\to\mathbb{R}\ \big|\ \vect{x}\text{ is measurable}\}.
\end{equation}

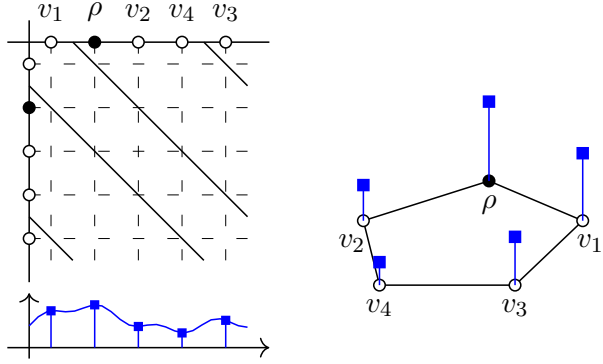
\begin{figure}
  \centering
  \resizebox{0.5\linewidth}{!}{\input{figs/graphing.tikz.tex}}
  \caption{
    Sampling of a random rooted graph from a graphing with an associated signal.
    \textbf{(Upper left)} A graphing $\mathbf{G}=(\hintCO{0,1},\set{E},\lambda)$.
    We sample from $\mathbf{G}$ by first selecting a root node $\rho$ at random (colored in black), and then taking the connected component of $\rho$ in the edge set.
    That is, starting with the node $\rho$, we see that the nodes $v_1,v_2$ are neighbors of $\rho$.
    Then, the nodes $v_3,v_4$ are neighbors of $v_1,v_2$, with the process stopping there.
    \textbf{(Lower left)} A graphing signal $\vect{x}:\hintCO{0,1}\to\mathbb{R}$. After sampling a node and edge set from the graphing, we sample this function at the corresponding values.
    \textbf{(Right)} The drawn rooted graph and signal sampled from the graphing.
  }
  \label{fig:graphing}
\end{figure}

Much like how a graphon describes a model for dense random graphs, a graphing carries with it a distribution of random rooted graphs.
As before, let $\nbhd$ be the neighborhood operator, returning all of the one-hop neighbors of a node $v\in\set{V}$, so that $\nbhd^K$ is the $K$-hop neighborhood operator.
For a graphing $\ging{g}=(\set{v},\set{e},\lambda)$ with associated signal $\vect{x}\in\mathbb{X}(\ging{g})$, the rooted $K$-ball at a node $r\in\set{V}$ is defined in the same way as in~\cref{sec:local-gsp}, denoted by $\rgra{G}_K(r)$, with the same notation used to denote the corresponding graph signal $\rvect{x}_K(r)$.
Due to the bounded degree condition on the graphing, $\rgra{G}_K(r)$ is always a finite rooted graph of maximum degree at most $D$.
With this in mind, we define a sampling map $\Sigma_K:\set{V}\to\Omega_K$ for graphings much like the sampling map~\eqref{eq:sampling-map} for finite graphs:
\begin{equation}
  \Sigma_K(v)=(\rgra{G}_K(v),\rvect{x}(v)).
\end{equation}
The sampling map $\Sigma_K$ is illustrated in \cref{fig:graphing} for a simple graphing.
Much like for finite graphs, we can pushforward the measure $\lambda$ to yield a probability measure $\mu$ on $\Omega_K$.
That is to say, $\mu=(\Sigma_K)_*(\ging{g},\vect{x})$, in the same way as before, where we adopt $\lambda$ as the probability measure over $\set{V}$ to be pushed forward.
This differs slightly from the case of finite graphs, where we implicitly assume that the initial measure was the uniform probability measure on the finite node set.

To establish the basic building blocks for signal processing on graphings, we first construct a Laplacian for graphing signals.
Unlike graph signals, graphing signals are not necessarily supported on a finite, or even countable space.
We specify the graphing Laplacian as a linear operator on the space of graphing signals in the following way.
For any $\vect{x}\in \mathbb{X}(\ging{g})$ and $v\in\set{V}$, put
\begin{equation}\label{eq:graphing-laplacian}
  [\mtx{S}\vect{x}]_{v} = \sum_{u\in\nbhd(v)}\vectidx{x}{v}-\vectidx{x}{u}.
\end{equation}
By the degree boundedness condition from \cref{defn:graphing}, the sum in \eqref{eq:graphing-laplacian} is well-defined.
Due to its singular nature, it is difficult to define a ``graphing Fourier transform'' directly from the Laplacian: for instance, even fairly tame graphing structures may have a Laplacian whose eigenvalues have uncountable multiplicity.
This, for instance, makes the notion of the power spectral distribution of a graphing signal unwieldy.

However, the spectral properties of a graphing signal with respect to the underlying graphing structure can be studied indirectly.
Let $\ging{g}$ be a graphing with graphing signal $\vect{x}\in\mathbb{X}(\ging{g})$ and Laplacian $\mtx{S}$.
Then, for integers $K\geq 0$, we tentatively define
\begin{equation}\label{eq:graphing-moments}
  \moment{m}_K(\vect{x}) = \int_\set{V} \vectidx{x}{v}[\mtx{S}^K\vect{x}]_v \drv\lambda(v),
\end{equation}
where $\mtx{S}^K$ again indicates $K$ repeated applications of the Laplacian.
See that this definition resembles that of the moments of the power spectral distribution of a graph signal in \eqref{eq:psd-moments}.
We define the values $\moment{m}_K(\vect{x})$ tentatively, as it is not obvious when they are finite, or even well-defined.
In the following discussion, we establish sufficient conditions under which the sequence $\{\moment{m}_K(\vect{x})\}_{K=0}^\infty$ exists and corresponds to the moments of a distribution function.

Much like in \cref{sec:spectra,sec:transfer}, we control the behavior of the graphing by bounding its size.
We first do this by determining a suitable notion of boundedness for a graphing signal.

\begin{definition}\label{defn:locally-essentially-bounded}
  Let $\ging{g}=(\set{v},\set{e},\lambda)$ be a graphing and $\vect{x}\in\mathbb{X}(\ging{g})$ be a graphing signal.
  The signal $\vect{x}$ is said to be locally essentially bounded if it is bounded on almost all of the connected components of $\ging{g}$.
  That is, there exists an $a\geq 0$ such that for all $K\geq 0$,
  \begin{equation}\label{eq:leb-condition}
    \lambda\left(\nbhd^K((\vect{x}^{-1}[-a,a])^{\comp})\right)=0,
  \end{equation}
  where $(\vect{x}^{-1}[-a,a])^{\comp}$ denotes the set of nodes $v\in\set{V}$ such that $\vectidx{x}{v}\notin[-a,a]$, and $\nbhd^K$ is the $K$-hop neighborhood of a set.
  Denote the set of all such graphing signals by $\mathbb{X}_{\LB}(\ging{g})$.
\end{definition}

In words, local essential boundedness not only controls the size of the set of nodes with large signal value (essential boundedness), but also controls the size of the neighborhoods with which those nodes can interact.
Local essential boundedness is an analog of boundedness for finite graph signals adapted to graphing signals.
For instance, a graphing signal that is strictly bounded is locally essentially bounded.
This condition, however, is stronger than being bounded almost everywhere (see~\cref{lemma:graphing-integrable}), due to the neighborhood condition in \eqref{eq:leb-condition}.
The necessity of this condition stems from the highly singular nature of the graphing Laplacian as an operator on signals.

We will also find it useful to consider graphings that ``resemble'' finite graphs in some sense.

\begin{definition}\label{defn:graph-convergence}
  Let $\ging{g}=(\set{v},\set{e},\lambda)$ be a graphing with graphing signal $\vect{x}\in\mathbb{X}(\ging{g})$, and $\{(\gra{g}_j,\vect{x}_j)\}_{j=1}^\infty$ a sequence of graphs with associated graph signals.
  For all $K\geq 0$, put $\mu_K=(\Sigma_K)_*(\ging{g},\vect{x})$, and $\mu_K(j)=(\Sigma_K)_*(\gra{g}_j,\vect{x}_j)$.
  Then, we say that the sequence of graphs $\{(\gra{g}_j,\vect{x}_j)\}_{j=1}^\infty$ converges weakly to $(\ging{g},\vect{x})$ if for all $K\geq 0$, $\mu_K(j)$ converges to $\mu_K$ as $j$ tends to infinity.
  We denote weak convergence of graphs to a graphing by $(\gra{g}_j,\vect{x}_j)\rightharpoonup(\ging{g},\vect{x})$.
\end{definition}

Much like \cref{thm:psd-convergence}, we can characterize the properties of a graphing based on the limiting properties of a sequence of graphs that converges to it.
We state this formally below.

\begin{theorem}\label{thm:graphing-psd}
  Let a graphing $\ging{g}$ of degree $D$ with signal $\vect{x}\in\mathbb{X}_{\LB}(\ging{G})$ be given.
  Suppose there exists a sequence of graphs and graph signals $\{(\gra{g}_j,\vect{x}_j)\}_{j=1}^\infty$ such that $(\gra{g}_j,\vect{x}_j)\rightharpoonup(\ging{g},\vect{x})$.
  Then, there exists a unique distribution function $P_{\vect{x}}$ supported on $[0,2D]$ such that the moments of $P_{\vect{x}}$ are given by the sequence $\{\mathbf{m}_K(\vect{x})\}_{K=0}^\infty$.
\end{theorem}

The proof of \cref{thm:graphing-psd} can be found in \cref{app:proof:graphing-psd}.
\Cref{thm:graphing-psd} defines the power spectral distribution for sufficiently nice graphing signals.
In particular, if the graphing and graphing signal are the limit of a sequence of graphs and graph signals, then asking that the power spectral distribution be continuous with respect to the weak topology on distributions forces the distribution to take on a unique value.
Namely, the power spectral distribution of the limit of a sequence of graphs and graph signals is merely the weak limit of the power spectral distributions of the graphs as defined in \eqref{eq:graph-psd}.
To show that this definition properly preserves the graph power spectral distribution of finite graphs, we illustrate with a simple example.

\begin{example}
  From a finite graph $\gra{g}=(\set{v},\set{e})$ on $n$ nodes and a graph signal $\vect{x}\in\mathbb{X}(\gra{g})$, a graphing $\ging{g}$ with graphing signal $\vect{y}\in\mathbb{X}_{\LB}(\ging{g})$ can be constructed in the following way~(\textit{cf.}~\cite[Example~18.16]{Lovasz:2012}).
  Identify $\set{v}$ with the integers $1,2,\ldots,n$.
  Let $\set{W}=\hintCO{0,1}$ with the usual Borel $\sigma$-field and the uniform (Lebesgue) probability measure $\lambda$.
  We construct a graphing $\ging{g}=(\set{W},\set{F},\lambda)$ in the following way.
  Partition $\set{W}$ into intervals $I_i=\hintCO{(i-1)/n,i/n}$ indexed by $i\in\set{v}$.
  Then, for each $(i,j)\in\set{e}$ with $i<j$, and every point $t\in I_i$, put the points $(t,t+(j-i)/n)\in I_i\times I_j, (t+(j-i)/n,t)\in I_j\times I_i$ into the edge set $\set{F}$.
  Additionally, assign $\vectidx{y}{t}=\vectidx{x}{i}$.

  Under this construction, one can show that for all $K\geq 0$, the distributions on $\Omega_K$ corresponding to $(\gra{g},\vect{x})$ and $(\ging{g},\vect{y})$ are equal, \ie, $(\Sigma_K)_*(\gra{g},\vect{x})=(\Sigma_K)_*(\ging{g},\vect{y})$.
  It holds, then, that $\moment{m}_K(\vect{x})=\moment{m}_K(\vect{y})$ for all $K$, so that the power spectral distribution corresponding to the graphing signal $\vect{y}$ is identical to that of the graph signal $\vect{x}$.
\end{example}

In~\cref{sec:spectra,sec:transfer,sec:weighted}, we considered how the distribution of rooted balls in graphs yields useful topological or metric structure with which graphs can be related to each other, particularly in light of graph summaries such as the power spectral distribution.
In all of these cases, the topology of weak convergence was used to show that a convergent sequence of graphs in the sense of the distribution of rooted balls is also convergent in the sense of any appropriate graph summary.
This left the question open: what do the graphs converge \emph{to}?
By considering graphings as limiting objects for sparse graphs, we have shown that the extension of graph summaries, such as the power spectral distribution, to graphings maintains continuity in the natural way.

\begin{remark}[Hyperfinite graphings]
  \Cref{thm:graphing-psd} hinges on the graphing $\ging{g}$ and signal $\vect{x}$ being the limit of a sequence of finite graphs and graph signals.
  As noted in \cite[Conjecture~19.8]{Lovasz:2012}, the question of whether all graphings are limits of sequences of graphs is still an open question.
  However, there is a class of graphings for which this is known to hold: namely, \emph{hyperfinite graphings}.
  We refer the interested reader to \cite[Chapter~21]{Lovasz:2012} for more details.
\end{remark}

\section{Conclusion}

Taking into account the folklore knowledge that low-order graph filters only use local information on graphs, we have developed a local framework in which many aspects of graph signal processing are illuminated.
In particular, by mapping a graph and graph signal to a probability distribution on a fixed space, we allow for the application of tools from probability theory to characterize similarities between graphs based on their local structures, even if they do not have a common node set.
Using the topology of weak convergence of probability distributions, we define a topology on the space of graphs and graph signals, where notions of continuity can be understood: this was most obviously demonstrated in \cref{sec:spectra}, where it was proved that the power spectral distribution of a graph signal is weakly continuous with respect to the distribution of rooted balls in a graph.
This approach allows for relatively painless treatment of limiting objects, such as \emph{graphings}, by simply decomposing them into their constituent distributions.

The analysis carried out in this paper provides proper justification for transferring graph filters across multiple graphs, given the assumption that they share local structural properties.
With the particular suitability of this approach to highly sparse graphs, this complements existing work on transferability based on graphon models or assumptions of graphs as discretizing manifold structures.
Leveraging the framework of this paper, it would be interesting to study more exotic types of functions, such as graph neural networks and graph kernels, or to incorporate the properties encoded by distributions of local structures into tasks such as network topology inference.

\appendix

\section{Metric Topology of the Signal Space}\label[app]{app:automorphism}

Let a rooted ball $\rgra{G}=(\set{V},\set{E},r)$ be given.
Denote by $\Aut(\rgra{G})$ the automorphism group of maps $\{\phi:\set{V}\to\set{V}\}$, \textit{i.e.}, the set of all isomorphisms from $\rgra{G}$ to itself, with the group operation given by composition.
The group $\Aut(\rgra{G})$ acts on $\mathbb{X}(\rgra{G})$ in a natural way.
For any $\phi\in\Aut(\rgra{G}), \vect{x}\in\mathbb{X}(\rgra{G})$, define the action of $\phi$ on $\vect{x}$ as the graph signal $\phi(\vect{x})$ satisfying
\begin{equation}
  [\phi(\vect{x})]_{\phi(v)} = \vectidx{x}{v}
\end{equation}
for all $v\in\set{V}$.
Next, define an equivalence relation on $\mathbb{X}(\rgra{G})$ where $\vect{x}\!\sim\!\phi(\vect{x})$ for all $\phi\in\Aut(\rgra{G}), \vect{x}\in\mathbb{X}(\rgra{G})$, and take the quotient space $\mathbb{X}(\rgra{G})/\!\sim$, whose elements are the equivalence classes under this relation.
Define the metric $\|\cdot\|_2$ on $\mathbb{X}(\rgra{G})/\!\sim$ so that for any $[\vect{x}],[\vect{y}]\in\mathbb{X}/\!\sim$,
\begin{equation}
  \|[\vect{x}]-[\vect{y}]\|_2 = \min_{\substack{\vect{x}'\in[\vect{x}]\\\vect{y}'\in[\vect{y}]}}\|\vect{x}'-\vect{y}'\|_2,
\end{equation}
where $\|\vect{x}'-\vect{y}'\|_2$ is the norm on $\mathbb{X}(\rgra{G})$ inherited from identification with $\mathbb{R}^{|\set{V}|}$.
As $\Aut(\rgra{G})$ is a finite group, this minimum is well-defined, and satisfies the triangle inequality, thus yielding a valid metric.
When defining the space $\Omega_K$ in \cref{sec:local-gsp:rdistro} we treat $\mathbb{X}(\rgra{G})$ with the structure of $\mathbb{X}(\rgra{G})/\!\sim$, from which the topology on $\Omega_K$ is inherited.
Similarly, we use the metric $\|\cdot\|_2$ in \cref{sec:transfer} as defined on $\mathbb{X}(\rgra{G})/\!\sim$.

\section{Proof of \texorpdfstring{\Cref{thm:psd-convergence}}{Theorem~\ref{thm:psd-convergence}}}\label[app]{app:proof:psd-convergence}

%
By \cref{lemma:compactness}, the sequence $\{\mu_j\}_{j=1}^\infty$ has compact support.
Since $\rmoment{m}_K$ is a continuous function, this implies that $\{\mathbb{E}_{\mu_j}[\rmoment{m}_K]\}_{j=1}^\infty$ is a convergent sequence by the definition of weak convergence of probability measures.
Applying \cref{prop:moments-expectations}, the moments $\moment{m}_K(\vect{x}_j)=\mathbb{E}_{\mu_j}[\rmoment{m}_K]$ converge, as desired.

For the case where weak convergence holds for all $K\geq 0$, we complete the proof with the following lemma.
\begin{lemma}\label{lemma:moment-uniqueness}
  Let $A=[0,a]$ be a compact subset of the real line, for some $a>0$.
  Let $\{\{\moment{m}_K(j)\}_{K=0}^\infty\}_{j=1}^\infty$ be an array of numbers such that $\{\moment{m}_K(j)\}_{K=0}^\infty$ is the sequence of moments of a finite measure $P_j$ supported on $A$, and that $\moment{m}_K(j)\overset{j}{\to}\moment{m}_K$ for each $K$, where $\{\moment{m}_K\}_{K=0}^\infty$ is some sequence of real numbers.

  Then, the sequence $\{\moment{m}_K\}_{K=0}^\infty$ uniquely corresponds to a distribution $P$ supported on $A$, such that the sequence of distributions $P_j$ converges weakly to $P$.
\end{lemma}
%
\Cref{lemma:moment-uniqueness} is a routine derivation following from the Hausdorff moment property~\cite[Theorem~3.15]{Schmudgen:2017} and the Weierstrass approximation theorem~\cite{Weierstrass:1885}, so we omit the proof.
In particular, let $A=[0,2D]$.
The moments $\moment{m}_K(\vect{x}_j)$ correspond to the power spectral distributions of the graph signals $\vect{x}_j$, which are supported on $A$ by the bounded degree assumption.
Since $\moment{m}_K(\vect{x}_j)$ converges for all $K\geq 0$, we have that the sequence of limits $\{\moment{m}_K\}_{K=0}^\infty$ is the sequence of moments of a distribution $P$ supported on $A$.
Moreover, we have that $P$ is the weak limit of the power spectral distributions of the graph signals $\vect{x}_n$, as desired.

\section{Proof of~\texorpdfstring{\Cref{lemma:compactness}}{Lemma~\ref{lemma:compactness}}}
\label[app]{app:proof:compactness}

Denote by $\rgraphsd{D}$ the set of rooted graphs such that the underlying graph has node degree bounded by $D$.
For any $K\geq 0$ and any graph $\gra{G}=(\set{v},\set{e})$ contained in $\graphsd{D}$, note that for all $v\in\set{v}$, it holds that the rooted $K$-ball centered at $v$ is contained in $\rgraphsd{D}$.
Moreover, it is straightforward to show that there are only finitely many rooted $K$-balls contained in $\rgraphsd{D}$, up to isomorphism.
For any $a\geq 0$ and any rooted graph $\rgra{G}$, denote the set of graph signals taking values in $[-a,a]$ by $\mathbb{X}(\rgra{G},a)$.
It is clear that $\mathbb{X}(\rgra{G},a)$ is compact as a subset of $\mathbb{X}(\rgra{G})$.

Using these constructions, it immediately follows that for any graph $\gra{g}\in\graphsd{D}$ with graph signal $\vect{x}\in\mathbb{X}(\rgra{G},a)$ for some $a>0$, the associated distribution $\mu=(\Sigma_K)_*(\gra{g},\vect{x})$ satisfies
\begin{equation}\label{eq:compact-support}
  \supp(\mu)\subseteq\coprod_{\rgra{G}_k\in\rgraphsd{D}:k\leq K}\mathbb{X}(\rgra{G},a).
\end{equation}
Since there are only finitely many rooted $K$-balls in $\rgraphsd{D}$, the above set is a disjoint union of finitely many compact spaces, and is thus a compact space itself, by the properties of the disjoint union of topological spaces.
If a family of graphs in $\{\gra{g}_j\}_{j\in J}\in\graphsd{D}$ has uniformly bounded signals $\{\vect{x}_j\}_{j\in J}$ for some index set $J$, this means that there is some $a\geq 0$ such that $\vect{x}_j\in\mathbb{X}(\gra{g}_j,a)$ for all $j\in J$, so that the condition in \eqref{eq:compact-support} holds for all graph/graph signals in the family, as desired.




\section{Proof of~\texorpdfstring{\Cref{thm:transferability}}{Theorem~\ref{thm:transferability}}}
\label[app]{app:proof:transferability}

We first establish that for a function $\rooted{J}$ satisfying \cref{assump:lipschitz}, the metric $d_C$ preserves Lipschitz continuity over the subspace $\Gamma\subset\Omega_K$ as defined in \cref{lemma:compactness}.
\begin{lemma}\label{lemma:lipschitz}
  Let $\rooted{J}:\Omega_K\to\mathbb{R}$ be a function satisfying \cref{assump:lipschitz}.
  Then, for any $C>0$, $\rooted{J}$ has a Lipschitz constant at most $\max\{L,A/C\}$ on the subspace $\Gamma$ with the metric $d_C$, where
  \begin{equation}
    A = \sup_{\omega\in\Gamma}\rooted{J}(\omega)-\inf_{\omega\in\Gamma}\rooted{J}(\omega).
  \end{equation}
\end{lemma}
\Cref{lemma:lipschitz} can be shown by directly applying the definition of Lipschitz continuity, so we omit the proof.
By the dual formulation of the $1$-Wasserstein distance~\cite{Sriperumbudur:2009}, \cref{lemma:lipschitz} implies that, for $C>0$,
\begin{equation}
  \big|\mathbb{E}_{\mu}[\rooted{J}]-\mathbb{E}_{\nu}[\rooted{J}]\big|
  \leq \max\left\{L,\frac{A}{C}\right\}\cdot W_1\left(\mu,\nu;C\right),
\end{equation}
owing to the fact that $\supp(\mu),\supp(\nu)\subseteq\Gamma$, by assumption.
Observing that $W_1(\mu,\nu;C)$ is monotonically increasing with respect to $C$, we can restrict our view to $C\geq A/L$.
Applying a simple change of variables yields the expression
\begin{equation}
  \big|\mathbb{E}_{\mu}[\rooted{J}]-\mathbb{E}_{\nu}[\rooted{J}]\big|
  \leq \inf_{C\in\hintOC{0,1}}\frac{L}{C}\cdot W_1\left(\mu,\nu;\frac{AC}{L}\right).
\end{equation}
Taking $C=1$ and noting that $A=1$ by assumption yields the bound~\eqref{eq:wasserstein-bound}, as desired.

\section{Proof of~\texorpdfstring{\Cref{thm:weighted-psd-convergence}}{Theorem~\ref{thm:weighted-psd-convergence}}}
\label[app]{app:proof:weighted-psd-convergence}

One can check that $\rmoment{m}_K$ is continuous and bounded on $\bigcup_{j=1}^\infty\supp(\mu_j)$ due to the weighted degree bound $D$ and the uniform boundedness of the signals, so that $\{\mathbb{E}_{\mu_j}[\rmoment{m}_K]\}_{j=1}^\infty$ is a convergent sequence by the definition of weak convergence of probability measures.
Applying \cref{prop:moments-expectations}, the moments $\moment{m}_K(\vect{x}_j)=\mathbb{E}_{\mu_j}[\rmoment{m}_K]$ converge, as desired.

When weak convergence holds for all $K\geq 0$, the proof is completed by appealing to \cref{lemma:moment-uniqueness} as in the proof of \cref{thm:psd-convergence}.

\section{Proof of~\texorpdfstring{\Cref{thm:graphing-psd}}{Theorem~\ref{thm:graphing-psd}}}
\label[app]{app:proof:graphing-psd}

We begin by establishing two results regarding locally essentially bounded signals $\mathbb{X}_{\LB}(\ging{g})$ and the graphing Laplacian $\mtx{S}$.

\begin{lemma}\label{lemma:graphing-integrable}
  For a graphing $\ging{g}=(\set{v},\set{e},\lambda)$, the set $\mathbb{X}_{\LB}(\ging{g})$ is a subset of $\mathcal{L}^{\infty}(\set{V},\lambda)$, where $\mathcal{L}^{\infty}(\set{V},\lambda)$ denotes the space of essentially bounded functions on $\set{V}$ with respect to the probability measure $\lambda$.
\end{lemma}
We omit the proof of \cref{lemma:graphing-integrable}, as it follows directly from \cref{defn:locally-essentially-bounded}.

\begin{lemma}\label{lemma:graphing-laplacian}
  $\mathbb{X}_{\LB}(\ging{g})$ is closed under application of the Laplacian $\mtx{S}$, \ie, for any $\vect{x}\in\mathbb{X}_{\LB}(\ging{g})$, we have that $\mtx{S}\vect{x}\in\mathbb{X}_{\LB}(\ging{g})$.
\end{lemma}
The proof of \cref{lemma:graphing-laplacian} can be found in \cref{app:proof:graphing-laplacian}.

For any $K\geq 0$, the graphing signal $\mtx{S}^K\vect{x}$ is contained in $\mathbb{X}_{\LB}(\ging{g})$ by \cref{lemma:graphing-laplacian}.
The product of two functions in $\mathcal{L}^\infty(\set{V},\lambda)$ is contained in $\mathcal{L}^\infty(\set{V},\lambda)$, so that the graphing signal $\vect{y}:v\to\vectidx{x}{v}\cdot[\mtx{S}^K\vect{x}]_v$ is contained in $\mathcal{L}^\infty(\set{V},\lambda)$, by \cref{lemma:graphing-integrable}.
Thus, $\moment{m}_K(\vect{x})=\mathbb{E}_\lambda[\vect{y}]$ has finite value for all $K\geq 0$.

Let $K\geq 0$ be given, and let $\rmoment{m}_K:\Omega_K\to\mathbb{R}$ be the function as defined in \cref{prop:moments-expectations}.
One can check that $\moment{m}_K(\vect{x})=\mathbb{E}_\mu[\rmoment{m}_K]$, where $\mu=(\Sigma_K)_*(\ging{g},\vect{x})$.
By the bounded degree property of the graphing and the assumption that $\vect{x}\in\mathcal{L}^\infty(\set{V},\lambda)$ (as a consequence of \cref{lemma:graphing-integrable}), \cref{lemma:compactness} implies that $\supp(\mu)$ is contained in a compact subset of $\Omega_K$.
For each finite graph and graph signal $(\gra{g}_n,\vect{x}_n)$, denote the $K^{th}$ moment of the signal $\vect{x}_n$ by $\moment{m}_K(\vect{x}_n)$, and the pushforward measure of $\Sigma_K$ on the space $\Omega_K$ by $\mu_n$, so that $\moment{m}_K(\vect{x}_n)=\mathbb{E}_{\mu_n}[\rmoment{m}_K]$, by \cref{prop:moments-expectations}.
The assumption that $(\gra{g}_n,\vect{x}_n)\rightharpoonup(\ging{g},\vect{x})$ implies that $\mu_n$ weakly converges to $\mu$.
Since $\mu$ has compact support and $\rmoment{m}_K$ is a continuous function, this implies that $\mathbb{E}_{\mu_n}[\rmoment{m}_K]\to\mathbb{E}_\mu[\rmoment{m}_K]$.
In other words, $\moment{m}_K(\vect{x}_n)\to\moment{m}_K(\vect{x})$.

Having satisfied the conditions of \cref{thm:psd-convergence}, we have that the power spectral distributions of the finite graphs converge to a unique distribution supported on $[0,2D]$, and that the moments of this distribution are given by $\{\moment{m}_K(\vect{x})\}_{K=0}^\infty$ by \cref{lemma:moment-uniqueness}, as desired.

\section{Proof of~\texorpdfstring{\Cref{lemma:graphing-laplacian}}{Lemma~\ref{lemma:graphing-laplacian}}}
\label[app]{app:proof:graphing-laplacian}

Let a degree $D$ graphing $\ging{G}=(\set{V},\set{E},\lambda)$ with graphing signal $\vect{x}\in\mathbb{X}_{\LB}(\ging{G})$ be given.
By definition, there exists an $a\geq 0$ such that for all $K\geq 0$, we have
\begin{equation}
  \lambda\left(\nbhd^K((\vect{x}^{-1}[-a,a])^{\comp})\right) = 0.
\end{equation}
Put $b=2Da$, and define the graphing signal $\vect{y}=\mtx{S}\vect{x}$, where $\mtx{S}$ denotes the graphing Laplacian.
Observe, due to the bounded degree $D$ of the graphing, that
\begin{equation}
  (\vect{y}^{-1}[-b,b])^{\comp}
  \subseteq
  \nbhd\left((\vect{x}^{-1}[-a,a])^{\comp}\right),
\end{equation}
so that, for all $K\geq 0$,
\begin{equation}\label{eq:nbhd-size-bound}
  \nbhd^K\left((\vect{y}^{-1}[-b,b])^{\comp}\right)
  \subseteq
  \nbhd^{K+1}\left((\vect{x}^{-1}[-a,a])^{\comp}\right).
\end{equation}
By the definition of local essential boundedness, the right hand side of~\eqref{eq:nbhd-size-bound} has measure zero (under the probability measure $\lambda$), which implies that the left hand side also has measure zero.
Since $K$ was arbitrary, this establishes that $\vect{y}\in\mathbb{X}_{\LB}(\ging{G})$.

\newpage
\bibliographystyle{IEEEtran}
\bibliography{myIEEEabrv,refs}

\end{document}

%% file: preamble.tex
\usepackage[T1]{fontenc}
\usepackage{amsmath,amsfonts,amssymb,amsthm,mathtools}
\usepackage{algorithm,algorithmic}

\usepackage{fullpage}

\usepackage[inline]{enumitem}

\usepackage{booktabs}

\usepackage{hyperref}
\hypersetup{
  colorlinks = true,
  allcolors = blue,
}
\usepackage[capitalise,nameinlink]{cleveref}
\usepackage{cite}
\usepackage{doi}

\crefformat{app}{#2Appendix #1#3}
\Crefformat{app}{#2Appendix #1#3}

\usepackage[x11names]{xcolor}

\usepackage{tikz}
\usetikzlibrary{cd}
\usetikzlibrary{shapes.geometric}
\usetikzlibrary{arrows.meta}
\usepgflibrary{plotmarks}

\usetikzlibrary{positioning}
\usetikzlibrary{backgrounds}
\usetikzlibrary{matrix}
\usetikzlibrary{decorations.markings}
\usetikzlibrary{calc}

\usepackage{pgfplots}
\pgfplotsset{compat=1.17}
\usepgfplotslibrary{groupplots}
\usepgfplotslibrary{fillbetween}

\pgfplotscreateplotcyclelist{mycyclelist}{%
  {red, mark=o, mark options={solid}, solid, thick},
  {blue, mark=x, mark options={solid}, dashed, thick},
  {violet, mark=*, mark options={solid}, loosely dotted, thick},
  {orange, mark=o, mark options={solid}, dashed, thick},
  {black, mark=x, mark options={solid}, solid, thick}
}
\pgfplotsset{cycle list name=mycyclelist}

\pgfplotscreateplotcyclelist{mygradlist}{%
{blue!20!white, mark=x, mark options={solid}, dotted, ultra thick},
{blue!40!white, mark=x, mark options={solid}, dashed, ultra thick},
{blue!60!white, mark=x, mark options={solid}, loosely dotted, ultra thick},
{blue!80!white, mark=x, mark options={solid}, loosely dashed, ultra thick},
{blue, mark=x, mark options={solid}, solid, ultra thick},
}

\newcounter{todo}

\makeatletter

\newcommand\listoftodos{\@starttoc{tod}}
\makeatother

\newtheorem{example}{Example}
\newtheorem{theorem}{Theorem}
\newtheorem{lemma}{Lemma}
\newtheorem{proposition}{Proposition}
\newtheorem{corollary}{Corollary}

\newtheorem{assumption}{Assumption}

\theoremstyle{definition}
\newtheorem{definition}{Definition}

\theoremstyle{remark}
\newtheorem{remark}{Remark}

\newenvironment{parremark}[1][]{\medskip\noindent\textbf{#1.}}{}
\newenvironment{parexample}[1][]{\medskip\noindent\textbf{Example: #1.}}{}

\newcommand*\xbar[1]{%
    \hbox{%
        \vbox{%
            \hrule height 0.66pt 
            \kern0.3ex
            \hbox{%
                \kern-0.1em
                \ensuremath{#1}%
                \kern-0.1em
            }%
        }%
    }%
}

\DeclareMathOperator{\Aut}{Aut}

\DeclareMathOperator{\supp}{supp}

\newcommand{\hintCO}[1]{\ensuremath{[#1)}}
\newcommand{\hintOC}[1]{\ensuremath{(#1]}}

\newcommand{\drv}{\ensuremath{\mathrm{d}}} 
\newcommand{\nbhd}{\ensuremath{\mathcal{N}}} 

\newcommand{\comp}{\ensuremath{\mathsf{C}}} 
\renewcommand{\top}{\ensuremath{\mathsf{T}}} 

\newcommand{\rooted}[1]{\ensuremath{\xbar{#1}}} 

\newcommand{\gra}[1]{\ensuremath{\mathcal{\uppercase{#1}}}} 
\newcommand{\rgra}[1]{\,\ensuremath{\rooted{\gra{#1}}}\,} 

\newcommand{\graphs}{\ensuremath{\mathbb{G}}} 
\newcommand{\graphsd}[1]{\ensuremath{\graphs_{#1}}} 
\newcommand{\rgraphsd}[1]{\ensuremath{\rooted{\graphs}_{#1}}} 
\newcommand{\graphsw}{\ensuremath{\graphs^W}} 
\newcommand{\graphswd}[1]{\ensuremath{\graphsw_{#1}}} 

\newcommand{\set}[1]{\ensuremath{\mathcal{\uppercase{#1}}}}
\newcommand{\rset}[1]{\ensuremath{\rooted{\set{#1}}}}

\newcommand{\ging}[1]{\ensuremath{\mathbf{\uppercase{#1}}}} 

\newcommand{\mtx}[1]{\ensuremath{\mathbf{\uppercase{#1}}}} 
\newcommand{\mtxidx}[2]{\ensuremath{{[\mtx{#1}]}_{#2}}} 
\newcommand{\rmtx}[1]{\,\ensuremath{\rooted{\mtx{#1}}}\,} 

\newcommand{\vect}[1]{\ensuremath{\mathbf{\lowercase{#1}}}} 
\newcommand{\vectidx}[2]{\ensuremath{[{\vect{#1}}]_{#2}}} 
\newcommand{\rvect}[1]{\,\ensuremath{\rooted{\vect{#1}}}\,} 
\newcommand{\rvectidx}[2]{\ensuremath{[{\rvect{#1}}]_{#2}}} 

\newcommand{\filt}[1]{\ensuremath{\mathcal{\uppercase{#1}}}} 
\newcommand{\rfilt}[1]{\,\ensuremath{\rooted{\filt{#1}}}\,}

\newcommand{\moment}[1]{\ensuremath{\mathbf{\lowercase{#1}}}} 
\newcommand{\rmoment}[1]{\ensuremath{\rooted{\mathbf{\lowercase{#1}}}}} 

\newcommand{\LB}{\ensuremath{\mathrm{LB}}} 

\newcommand{\ie}{\textit{i.e.}}

\newcommand{\iid}{\textnormal{i.i.d.}}

%% file: figs/motifspace.tikz.tex
\begin{tikzpicture}
  \tikzset{
    bigcirc/.style={
      circle,
      inner sep=0pt,
      text width=5mm,
      align=center,
      draw=black,
    }
  }

  \tikzset{
    smallcirc/.style={
      circle,
      inner sep=0pt,
      text width=3mm,
      align=center,
      draw=black,
    }
  }
  
  \draw[<->] (5,0) -- (17,0);

  \begin{scope}[on background layer]
    \draw[draw=none, fill=white!95!black] (4,3) rectangle (18,-2);
    \node[below left] at (18,3) {\Large$\Omega_1$};
  \end{scope}

  
  \foreach \x/\idx/\dim in {5.75/1/2, 8.25/2/3, 10.75/3/3, 13.25/4/4} {
    \draw[draw=black,fill=white] ($(\x,2)+(0.1,-0.1)$) rectangle ($(\x,0)+(2.4,0.1)$);
    \node[above] at ($(\x,2)+(1.2,0)$) {$\mathbb{X}(\rgra{G}_{\idx})\cong\mathbb{R}^{\dim}$};
  }
  \node at (5,1) {$\dots$};
  \node at (17,1) {$\dots$};

  
  \foreach \l/\x/\y/\color in {A_1/6.5/-0.5/black, A_2/7.5/-1.0/white,
    B_1/9.5/-0.5/black, B_2/9/-1.0/white, B_3/10/-1.0/white,
    C_1/11.5/-1.0/black, C_2/12/-0.5/white, C_3/12.5/-1.0/white,
    D_1/14/-1.0/black, D_2/14/-0.5/white, D_3/15/-0.5/white, D_4/15/-1.0/white} {
    \node (\l) at (\x,\y) [smallcirc,fill=\color,draw] {};
  }

  \foreach \i/\j in {A_1/A_2,
    B_1/B_2, B_1/B_3,
    C_1/C_2, C_1/C_3, C_2/C_3,
    D_1/D_2, D_1/D_3, D_1/D_4} {
    \path (\i) edge (\j);
  }

  \node at (5,-1) {$\dots$};
  \node at (17,-1) {$\dots$};

  \foreach \x/\idx in {7/1, 9.5/2, 12/3, 14.5/4} {
    \node[below] at (\x,-1) {$\rgra{G}_{\idx}$};
  }

  
  \pgfmathsetseed{1883}
  
  \foreach \l/\x/\y in {a/0/1, d/2/0, g/3/2, h/3/0} {
    \pgfmathsetmacro{\mix}{100*(rand+1)/2}
    \def\color{blue!\mix!white!50!white}

    \node (\l) at (\x,\y) [bigcirc,draw,fill=\color] {$\mathsf{\l}$};

    \pgfmathsetmacro{\xcoord}{(rand+1)*2.0/2+5.75+0.2}
    \pgfmathsetmacro{\ycoord}{(rand+1)*1.2/2+0.4}
    \node at (\xcoord,\ycoord) [smallcirc,draw,fill=\color] {\small{$\mathsf{\l}$}};
  }
  
  \foreach \l/\x/\y in {b/1/0, e/2/2} {
    \pgfmathsetmacro{\mix}{100*(rand+1)/2}
    \def\color{blue!\mix!white!50!white}

    \node (\l) at (\x,\y) [bigcirc,draw,fill=\color] {$\mathsf{\l}$};

    \pgfmathsetmacro{\xcoord}{(rand+1)*2.0/2+8.25+0.2}
    \pgfmathsetmacro{\ycoord}{(rand+1)*1.2/2+0.4}
    \node at (\xcoord,\ycoord) [smallcirc,draw,fill=\color] {\small{$\mathsf{\l}$}};
  }
  
  \foreach \l/\x/\y in {c/1/1, f/2/1} {
    \pgfmathsetmacro{\mix}{100*(rand+1)/2}
    \def\color{blue!\mix!white!50!white}

    \node (\l) at (\x,\y) [bigcirc,draw,fill=\color] {$\mathsf{\l}$};

    \pgfmathsetmacro{\xcoord}{(rand+1)*2.0/2+13.25+0.2}
    \pgfmathsetmacro{\ycoord}{(rand+1)*1.2/2+0.4}
    \node at (\xcoord,\ycoord) [smallcirc,draw,fill=\color] {\small{$\mathsf{\l}$}};
  }

  
  \foreach \i/\j in {a/b, b/c, c/d, c/e, e/f, f/g, f/h} {
    \path (\i) edge (\j);
  }

  
  \draw[->, below] (1.5,-0.4) to[bend right, left] node[midway, below, inner sep=4pt] {\large$\Sigma_1$} (3.8,-1.2);
\end{tikzpicture}

%% file: figs/wasserstein.tikz.tex
\begin{tikzpicture}
  %
  %

  \begin{scope}[shift={(0,1)}]
    \foreach \l/\x/\y in {0/0/0, 00/-0.5/-1, 01/0.5/-1, 000/-1/-2, 010/0/-2, 011/1/-2} {
      \pgfmathsetmacro{\mix}{100*(rand+1)/2}
      \node (a\l) at (\x,\y) [shape=circle,draw,fill=blue!\mix!white!50!white] {};
    }
    
    \foreach \i/\j in {a0/a00, a0/a01, a00/a000, a01/a010, a01/a011} {
      \path (\i) edge (\j);
    }
    
    \node at (0,-2.5) {(a)};
  \end{scope}

  \begin{scope}[shift={(3,0)}]
    \pgfmathsetmacro{\mix}{100*(rand+1)/2}
    \node (b0) at (0,0) [shape=circle,draw,fill=blue!\mix!white!50!white] {};
    \foreach \l/\t in {1/30, 2/90, 3/150, 4/210, 5/270, 6/330} {
      \pgfmathsetmacro{\mix}{100*(rand+1)/2}
      \node (b\l) at (\t:1) [shape=circle,draw,fill=blue!\mix!white!50!white] {};
      \path (b0) edge (b\l);
    }
    
    \foreach \i/\j in {b1/b2, b2/b3, b3/b4, b4/b5} {
      \path (\i) edge (\j);
    }
    \path[color=white, line width=3pt] (b5) edge (b1);
    \path (b5) edge (b1);
    
    \node at (0,-1.5) {(b)};
  \end{scope}

  \begin{scope}[shift={(6,0)}]
    \pgfmathsetmacro{\mix}{100*(rand+1)/2}
    \node (c0) at (0,0) [shape=circle,draw,fill=blue!\mix!white!50!white] {};
    \foreach \l/\t in {1/60, 2/120, 3/180, 4/240, 5/300, 6/360} {
      \pgfmathsetmacro{\mix}{100*(rand+1)/2}
      \node (c\l) at (\t:1) [shape=circle,draw,fill=blue!\mix!white!50!white] {};
      \path (c0) edge (c\l);
    }
    
    \foreach \i/\j in {c1/c2, c2/c3, c3/c4, c5/c6, c6/c1} {
      \path (\i) edge (\j);
    }
    
    \node at (0,-1.5) {(c)};
  \end{scope}

  \begin{scope}[shift={(-1.0,-6.25)}, xscale=0.9]
    \foreach \y in {3.0,1.5,0.0} {
      \draw[->] (0,\y) -- (8.2,\y);
      \node[right] at (8.2,\y) {$\Omega_1$};
      \draw[->] (0,\y) -- (0,\y+1.25);
    }
    
    \node[left] at (0,3.75) {(a)};
    \draw[color=blue, draw opacity=0.2, line width=24pt] plot[ycomb, yshift=3.0cm, yscale=1.75]
    coordinates {(4/7, 0.5) (12/7, 0.33) (20/7, 0.167) (28/7, 0) (36/7, 0) (44/7, 0) (52/7, 0)};

    \node[left] at (0,2.25) {(b)};
    \draw[color=blue, draw opacity=0.2, line width=24pt] plot[ycomb, yshift=1.5cm, yscale=1.75]
    coordinates {(4/7, 0.143) (12/7, 0) (20/7, 0) (28/7, 0) (36/7, 0.143) (44/7, 0) (52/7, 0.714)};

    \node[left] at (0,0.75) {(c)};
    \draw[color=blue, draw opacity=0.2, line width=24pt] plot[ycomb, yscale=1.75]
    coordinates {(4/7, 0) (12/7, 0) (20/7, 0) (28/7, 0.256) (36/7, 0) (44/7, 0.143) (52/7, 0.571)};
  \end{scope}

  \begin{scope}[shift={(-1,-6.5)}, every node/.style={scale=0.5, shape=circle, draw}, xscale=0.9]
    \node (m1_r) at (4/7, 0) [fill=black] {};
    \node (m1_l) at (4/7, -0.5) {};
    \path (m1_r) edge (m1_l);

    \node (m2_r) at (12/7, 0) [fill=black] {};
    \node (m2_l1) at (10/7, -0.5) {};
    \node (m2_l2) at (14/7, -0.5) {};
    \path (m2_r) edge (m2_l1)
    (m2_r) edge (m2_l2);

    \node (m3_r) at (20/7, -0.3) [fill=black] {};
    \node (m3_l1) at (18/7, -0.5) {};
    \node (m3_l2) at (20/7, 0) {};
    \node (m3_l3) at (22/7, -0.5) {};
    \path (m3_r) edge (m3_l1)
    (m3_r) edge (m3_l2)
    (m3_r) edge (m3_l3);

    \node (m4_r) at (28/7, 0) [fill=black] {};
    \node (m4_l1) at (26/7, -0.5) {};
    \node (m4_l2) at (30/7, -0.5) {};
    \path (m4_r) edge (m4_l1)
    (m4_r) edge (m4_l2)
    (m4_l1) edge (m4_l2);

    \node (m5_r) at (36/7, -0.25) [fill=black] {};
    \node (m5_l1) at (34/7, 0) {};
    \node (m5_l2) at (36/7, 0) {};
    \node (m5_l3) at (38/7, 0) {};
    \node (m5_l4) at (38/7, -0.5) {};
    \node (m5_l5) at (36/7, -0.5) {};
    \node (m5_l6) at (34/7, -0.5) {};
    \path (m5_r) edge (m5_l1)
    (m5_r) edge (m5_l2)
    (m5_r) edge (m5_l3)
    (m5_r) edge (m5_l4)
    (m5_r) edge (m5_l5)
    (m5_r) edge (m5_l6);
    \draw (m5_l1) -- (m5_l2) -- (m5_l3) -- (m5_l5) -- (m5_l6) -- (m5_l1);

    \node (m6_r) at (44/7, -0.25) [fill=black] {};
    \node (m6_l1) at (42/7, 0) {};
    \node (m6_l2) at (44/7, 0) {};
    \node (m6_l3) at (46/7, 0) {};
    \node (m6_l4) at (46/7, -0.5) {};
    \node (m6_l5) at (44/7, -0.5) {};
    \node (m6_l6) at (42/7, -0.5) {};
    \path (m6_r) edge (m6_l1)
    (m6_r) edge (m6_l2)
    (m6_r) edge (m6_l3)
    (m6_r) edge (m6_l4)
    (m6_r) edge (m6_l5)
    (m6_r) edge (m6_l6);
    \draw (m6_l1) -- (m6_l2) -- (m6_l3) -- (m6_l4) -- (m6_l5) -- (m6_l6);

    \node (m7_r) at (54/7, 0) [fill=black] {};
    \node (m7_l1) at (50/7, 0) {};
    \node (m7_l2) at (50/7, -0.5) {};
    \node (m7_l3) at (54/7, -0.5) {};
    \path (m7_r) edge (m7_l1)
    (m7_r) edge (m7_l2)
    (m7_r) edge (m7_l3);
    \draw (m7_l1) -- (m7_l2) -- (m7_l3);
  \end{scope}
  
\end{tikzpicture}

%% file: figs/graphing.tikz.tex
\begin{tikzpicture}[scale=2, domain=0:1, every node/.style={scale=0.7}]
  \draw[thin] (-0.1,0) -- (1.1,0);
  \draw[thin] (0,0.1) -- (0,-1.1);

  \draw (0.2,0) -- (1.0,-0.8);
  \draw (0,-0.2) -- (0.8,-1.0);
  \draw (0.8,0) -- (1.0,-0.2);
  \draw (0,-0.8) -- (0.2,-1.0);


  \draw[very thin, loosely dashed] (0,-0.3) -- (1.0,-0.3);
  \draw[very thin, loosely dashed] (0.3,0) -- (0.3,-1.0);
  \draw[fill=black] (0,-0.3) circle (0.75pt);
  \draw[fill=black] (0.3,0) circle (0.75pt);
  \node[anchor=south] at (0.3,0.05) {$\rho$};
  
  \draw[very thin, loosely dashed] (0,-0.1) -- (1.0,-0.1);
  \draw[very thin, loosely dashed] (0.1,0) -- (0.1,-1.0);
  \draw[very thin, loosely dashed] (0,-0.5) -- (1.0,-0.5);
  \draw[very thin, loosely dashed] (0.5,0) -- (0.5,-1.0);
  \draw[fill=white] (0,-0.1) circle (0.75pt);
  \draw[fill=white] (0.1,0) circle (0.75pt);
  \node[anchor=south] at (0.1,0.05) {$v_1$};
  \draw[fill=white] (0,-0.5) circle (0.75pt);
  \draw[fill=white] (0.5,0) circle (0.75pt);
  \node[anchor=south] at (0.5,0.05) {$v_2$};
  
  \draw[very thin, loosely dashed] (0,-0.9) -- (1.0,-0.9);
  \draw[very thin, loosely dashed] (0.9,0) -- (0.9,-1.0);
  \draw[very thin, loosely dashed] (0,-0.7) -- (1.0,-0.7);
  \draw[very thin, loosely dashed] (0.7,0) -- (0.7,-1.0);
  \draw[fill=white] (0,-0.9) circle (0.75pt);
  \draw[fill=white] (0.9,0) circle (0.75pt);
  \node[anchor=south] at (0.9,0.05) {$v_3$};
  \draw[fill=white] (0,-0.7) circle (0.75pt);
  \draw[fill=white] (0.7,0) circle (0.75pt);
  \node[anchor=south] at (0.7,0.05) {$v_4$};

  \begin{scope}[yshift=-1.4cm]
    \draw[->, thin] (0,-0.05) -- (0,0.25);
    \draw[->, thin] (-0.1,0) -- (1.1,0);
    \draw[color=blue, thin] plot
    (\x,{0.1+0.05*sin(4*\x r)+0.05*sin(8*\x r)+0.02*sin(24*\x r)});
    \draw[color=blue] plot[ycomb, mark=square*, mark size=0.5pt, samples=5, domain=0.1:0.9]
    (\x,{0.1+0.05*sin(4*\x r)+0.05*sin(8*\x r)+0.02*sin(24*\x r)});
  \end{scope}

  \def\ngon{5}
  \node[xslant=0.4,yscale=0.5,regular polygon,regular polygon sides=\ngon,minimum size=3cm] (p) at (2.0,-0.9) {};
  \draw[thin] (p.corner 1) -- (p.corner 2) -- (p.corner 3) -- (p.corner 4) -- (p.corner 5) -- cycle;
  \draw[fill=black] (p.corner 1) circle (0.75pt);
  \draw[fill=white] (p.corner 2) circle (0.75pt);
  \draw[fill=white] (p.corner 3) circle (0.75pt);
  \draw[fill=white] (p.corner 4) circle (0.75pt);
  \draw[fill=white] (p.corner 5) circle (0.75pt);

  \node[yshift=-0.05cm, anchor=north] at (p.corner 1) {$\rho$};
  \node[yshift=-0.05cm, xshift=-0.15cm, anchor=north]
  at (p.corner 2) {$v_2$};
  \node[yshift=-0.05cm, anchor=north] at (p.corner 3) {$v_4$};
  \node[yshift=-0.05cm, anchor=north] at (p.corner 4) {$v_3$};
  \node[yshift=-0.05cm, xshift=0.1cm, anchor=north]
  at (p.corner 5) {$v_1$};

  \begin{scope}[yscale=2]
    \draw[-{Square}, color=blue] (p.corner 1) -- ++(0,0.196cm);
    \draw[-{Square}, color=blue] (p.corner 2) -- ++(0,0.096cm);
    \draw[-{Square}, color=blue] (p.corner 3) -- ++(0,0.067cm);
    \draw[-{Square}, color=blue] (p.corner 4) -- ++(0,0.125cm);
    \draw[-{Square}, color=blue] (p.corner 5) -- ++(0,0.169cm);
  \end{scope}

\end{tikzpicture}